# GeoJSON Agents: A Multi-Agent LLM Architecture for Geospatial Analysis — Function Calling vs Code Generation

Qianqian Luo[a,†], Qingming Lin[a,†], Liuchang Xu[a,b,*], Sensen Wu[b], Ruichen Mao[c], Chao Wang[b], Hailin Feng[a], Bo Huang[d] and Zhenhong Du[b]

**a.* School of Mathematics and Computer Science, Zhejiang Agriculture and Forestry University, Hangzhou, China;*

*b. School of Earth Sciences, Zhejiang University, Hangzhou, China;*

*c. Moganshan Geospatial Information Laboratory, Huzhou, China;*

*d. Department of Geography, The University of HongKong, Hong Kong, China*

*Corresponding author: Liuchang Xu, email：xuliuchang@zafu.edu.cn

†: These authors contributed equally to this work and should be considered co-first authors.

# GeoJSON Agents: A Multi-Agent LLM Architecture for Geospatial Analysis — Function Calling vs Code Generation

## Abstract

Large Language Models (LLMs) have demonstrated substantial progress in task automation and natural language understanding. However, without domain expertise in geographic information science (GIS), they continue to encounter limitations including reduced accuracy and unstable performance when processing complex spatial tasks. To address these challenges, we propose GeoJSON Agents—a novel multi-agent LLM architecture specifically designed for geospatial analysis. This framework transforms natural language instructions into structured GeoJSON operations through two widely adopted LLM enhancement techniques: Function Calling and Code Generation. The architecture integrates three core components: task parsing, agent collaboration, and result integration. The Planner agent systematically decomposes user-defined tasks into executable subtasks, while specialized Worker agents perform spatial data processing and analysis either by invoking predefined function APIs or by dynamically generating and executing Python-based analytical code. The system produces reusable, standards-compliant GeoJSON outputs through iterative refinement. To systematically evaluate both approaches, we constructed a hierarchical benchmark comprising 70 tasks spanning basic, intermediate, and advanced complexity levels, conducting experiments with OpenAI's GPT-4o as the core model. Results indicate that the Code Generation–based agent achieved 97.14% accuracy, while the Function Calling–based agent attained 85.71%—both significantly outperforming the best-performing general-purpose model (48.57%). Comparative analysis reveals that Code Generation offers superior flexibility for complex, open-ended tasks, whereas Function Calling provides enhanced execution stability for structured operations. This study represents the first systematic integration of GeoJSON data with a multi-agent LLM framework and provides empirical evidence comparing two mainstream enhancement methodologies in geospatial contexts, offering new perspectives for improving GeoAI system performance and reducing barriers to GIS application.

## 1. Introduction

With the explosive growth of geospatial data and the increasing demand for spatial intelligence, traditional geographic information systems (GIS) are facing an urgent need to evolve from conventional software tools into intelligent systems(Maguire & D. J, 1991; Goodchild, 2005). However, in practical applications, traditional GIS tools typically require users to possess specialized domain knowledge and technical skills, creating a high entry barrier. Complex tasks often depend on the coordination of multiple tools and manual scripting, which makes execution time-consuming, error-prone, and labor-intensive, falling short of meeting the growing demands of spatial analysis(Yeh & A. G, 1999; Kahila-Tani et al., 2019; Khashoggi & Murad, 2020; Singh, 2019). To overcome these limitations, geospatial artificial intelligence (GeoAI) has emerged. It integrates artificial intelligence with geospatial computing to enable the automation and intelligent execution of spatial tasks(Hochmair et al., 2025). In particular, recent advances in LLMs—especially in natural language understanding, code generation, and knowledge integration(X. Huang et al., 2025)—have opened new opportunities for the development of autonomous GIS systems, where users can execute and plan tasks through natural language instructions. However, most mainstream LLMs are general-purpose models that lack a deep understanding of domain-specific geospatial knowledge, including spatial logic, topological relationships, and geospatial data structures. As a result, they often produce inaccurate outputs and generate incoherent steps when applied to spatial tasks(Mooney et al., 2023; L. Xu et al., 2025). These limitations become more pronounced when tasks involve multi-tool coordination and dynamic task planning, where general LLMs often show poor stability and limited controllability—failing to meet the accuracy and reliability standards required in professional GIS workflows(Wei et al., 2022).

To overcome the inherent limitations of general-purpose LLMs and enhance their performance in complex tasks, researchers have introduced LLM-driven agent frameworks to enable automated task planning and efficient execution. These frameworks typically consist of one or more agents equipped with reasoning and action capabilities, allowing them to dynamically perceive their environment, perform tasks, and provide feedback. This architecture significantly improves the reliability and stability of LLMs in complex geospatial analysis tasks(Lin et al., 2024; C. Wang et al., 2025). Currently, there are two primary execution strategies for driving agents: function calling and code generation. In June 2023, OpenAI introduced the concept of "function calling," whereby an LLM is provided with a set of predefined function interfaces that allow it to invoke external functions or services. This mechanism enriches the LLM's contextual understanding and enables it to produce highly structured outputs. This method has been widely adopted in domains such as financial research(Otiefy & Alhamzeh, 2024) and healthcare informatics(Z. Huang et al., 2024). For example, the Wolfram Alpha ChatGPT plugin(Wolfram, 2024) leverages function calling to perform complex computational queries and data retrieval tasks(M. Wang et al., 2025). However, function calling is heavily dependent on the completeness of the function library, making it less adaptable to new or evolving tasks. In contrast, the code generation approach enables agents to generate customized code based on natural language descriptions, leveraging specialized libraries (e.g., GeoPandas, Shapely) to perform complex, non-standardized geospatial analyses. This method offers greater flexibility and generalization capability. To date, LLMs such as AlphaCode(Y. Li et al., 2022), CodeGen(Nijkamp, Pang, et al., 2023), ChatGPT(OpenAI, 2023a), and CodeLama(Rozière et al., 2024) have demonstrated effective code generation capabilities. This approach has been widely adopted in software development; for instance, systems like Copilot(M. Chen et al., 2021) have significantly improved developer productivity(Dong et al., 2025). However, it also introduces challenges such as high execution uncertainty and reduced system stability. In the field of GeoAI, both methods have undergone preliminary exploration. For instance, in ShapefileGPT(Lin et al., 2024), the function calling approach successfully processed 42 shapefile-based spatial analysis tasks with an accuracy of 92.86%. In another study, a pre-trained model, GeoCode-GPT, was proposed to focus specifically on geospatial code generation(Hou et al., 2025). Both function calling and code generation offer novel pathways for enabling the automated execution of geospatial tasks using LLMs.

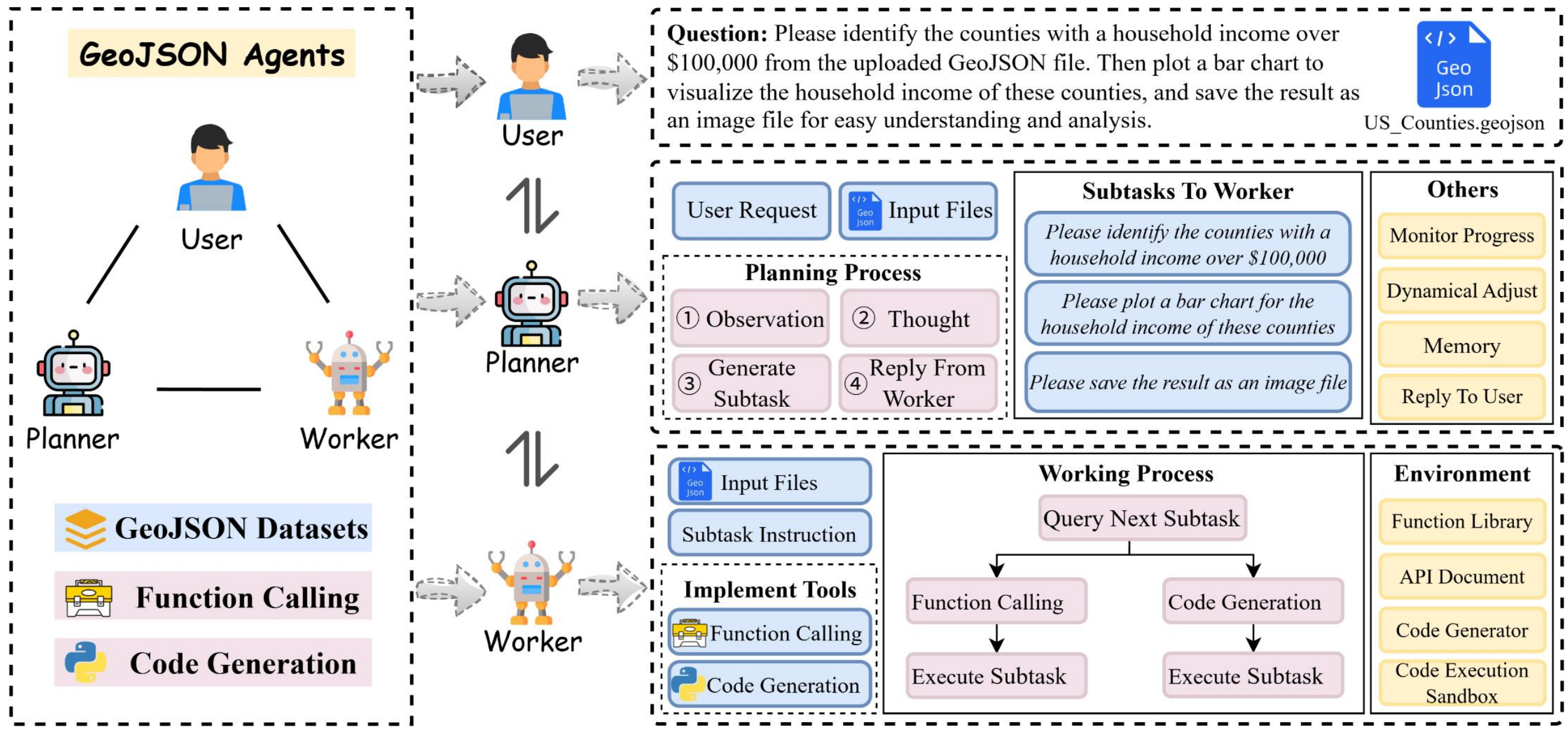

**Figure 1. Overall architecture of our GeoJSON Agents.** Both the Planner and Worker are LLM-powered agents that communicate through structured prompts in a closed-loop feedback system. The user uploads spatial task instructions and input files. The Planner agent converts ambiguous natural language instructions into decomposed subtasks and assigns them to the Worker agent. Upon receiving commands from the Planner, the Worker selects and invokes the most appropriate tools to execute assigned tasks and returns results to the Planner for iterative refinement until task completion.

Geospatial analysis tasks are highly dependent on the choice of spatial data format, as different vector data formats vary significantly in terms of structural design, storage methods, and suitable application scenarios. Commonly used vector data formats include Shapefile, GeoPackage, and GeoJSON. Among them, Shapefile is one of the most classic and widely used formats in the GIS domain. It features a simple and intuitive vector data storage structure, making it well-suited for desktop GIS analysis tasks(Lin et al., 2024; Pantiukhin et al., 2025). However, Shapefile has several limitations, including a fragmented multi-file structure, binary encoding that hinders direct parsing, and poor compatibility with modern WebGIS platforms. These drawbacks limit its effectiveness in complex analytical tasks and web-based applications. GeoPackage, a newer spatial data format proposed by the Open Geospatial Consortium (OGC), is based on the SQLite database engine and supports the integration of multiple data types within a single file. It offers high data integrity and efficient querying. However, like Shapefile, its binary encapsulation limits ease of direct interaction and parsing(Pantiukhin et al., 2025). In comparison, GeoJSON is a lightweight, open-source data format based on JSON. Its clear structure and ease of parsing and generation make it well-suited for data exchange between modern WebGIS applications and artificial intelligence models. As such, it is rapidly becoming a new standard for spatial data exchange(Pantiukhin et al., 2025). Recent studies have attempted to integrate traditional Shapefile formats with LLMs. For instance, Lin et al., (2024) proposed ShapefileGPT, which used function calling to handle simple spatial analysis tasks stored in Shapefile format, achieving promising initial results. However, the inherent limitations of the Shapefile format constrain its further development in LLM-driven complex geospatial task processing. Therefore, this study adopts the more modern and LLM-compatible GeoJSON format to explore its deep integration with an LLM-based multi-agent architecture, offering a new direction for achieving more complex and intelligent geospatial analysis tasks.

Therefore, this study aims to explore the integration of GeoJSON with an LLM-based multi-agent architecture and to evaluate the effectiveness of function calling and code generation in geospatial tasks. To this end, we propose and implement two LLM-driven multi-agent models, named GeoJSON Agents. These models adopt function calling and code generation approaches, respectively, and are specifically designed to handle spatial analysis tasks in the GeoJSON format. Both GeoJSON Agents share a unified architecture composed of two key components: a Planner and a Worker. The Planner interprets user intent and decomposes tasks, while the Worker executes them through predefined function calls or dynamically generated Python code, enabling automated geospatial analysis driven by natural language instructions. We also constructed a GeoJSON task benchmark encompassing three levels of complexity—basic, intermediate, and advanced—along with corresponding real-world datasets. This benchmark was used to evaluate the performance of the two GeoJSON Agents on GeoJSON-based spatial analysis tasks and to compare their results against those of a general-purpose LLM. In addition, we investigated the performance differences between the two GeoJSON Agents—based on function calling and code generation—when handling geospatial tasks of varying complexity. In our experiments, the function-calling-based GeoJSON Agent achieved a task accuracy of 85.71%, while the code-generation-based agent reached 97.14%. Both significantly outperformed the general-purpose model, demonstrating the superior performance of GeoJSON Agents in geospatial analysis tasks. Moreover, we observed that the code generation approach offers greater flexibility in handling complex, high-degree-of-freedom tasks, whereas the function calling approach exhibits better execution stability and efficiency. These findings provide a reliable basis for selecting appropriate methods across different application scenarios.

The significance of this study lies not only in addressing the limitations of traditional GIS workflows and enabling efficient, automated processing of geospatial tasks, but also in effectively overcoming the accuracy and controllability challenges faced by general-purpose models due to the lack of domain-specific geographic knowledge. This work is the first to integrate GeoJSON—a modern and lightweight spatial data format—into an LLM-based multi-agent framework, and to develop two multi-agent models employing distinct execution strategies: function calling and code generation. These models are designed for automated spatial analysis, filling a notable research gap in the field. Furthermore, the systematic comparison of these two implementation methods across varying task complexities provides quantitative evidence and theoretical guidance for future researchers in selecting suitable agent enhancement strategies for different application scenarios. Additionally, the proposed model architecture and task benchmark framework offer a novel paradigm for geospatial analysis and contribute to the advancement of autonomous GIS systems for geospatial tasks. The specific contributions of this study are as follows:

(1) We designed and implemented two LLM-driven multi-agent models—GeoJSON Agents—based on function calling and code generation, respectively. By integrating GeoJSON vector data, these models lower the barrier to using GIS tools, address the performance limitations of general-purpose models, and establish a unified, efficient, and automated mechanism for executing geospatial tasks.

(2) We developed a GeoJSON task benchmark covering three levels of task complexity, accompanied by real-world geographic datasets, and systematically evaluated the performance of the GeoJSON Agents.
(3) We conducted a systematic comparison of the two GeoJSON Agents based on their respective implementation methods, analyzed their performance across tasks of varying complexity, and provided quantitative insights and design references for future model development and method selection.

## 2. Related Work

### *2.1. LLM Agents in GeoAI*

The rapid advancement of large language models (LLMs) has given rise to the emergence of intelligent agents. Conceptually, an agent can be decomposed into three fundamental components: the brain, which serves as the center of cognition and decision-making; perception, which handles information input; and action, which executes tasks(L. Wang et al., 2024). Based on this core framework, agent architectures can be categorized into single-agent and multi-agent systems(Hong et al., 2023). Research on single-agent systems focuses on continuously enhancing the autonomy and problem-solving capabilities of an individual agent. To improve single-agent performance, the pioneering ReAct framework(Yao et al., 2023) was the first to integrate "reasoning" and "acting" into a dynamic loop, surpassing conventional static prompting methods. Building on this, the RAISE framework(Oh et al., 2025) introduced a memory mechanism to address the issue of context forgetting in long task chains. Self-Rewarding Language Models(Yuan et al., 2025) further advanced this approach by incorporating an LLM-as-a-Judge mechanism and an iterative training framework (Iterative DPO), enabling agents to evaluate and score their own responses. This established a self-rewarding system that moves beyond traditional human preference feedback. Despite these advancements, single-agent architectures still exhibit inherent limitations. When dealing with intrinsically complex, multi-step tasks, prolonged reasoning processes can lead to hallucinations, and the inability to process tasks in parallel constrains their reliability and efficiency in complex application scenarios. In contrast, multi-agent systems comprise multiple agents with distinct roles, each possessing different capabilities and behavioral strategies. These agents can observe their environment, collaborate, decompose tasks, and solve complex problems collectively(C. Wang et al., 2025). Several advanced autonomous systems have already employed LLM-driven multi-agent architectures to handle complex tasks(Z. Li & Ning, 2023; Ning et al., 2025). Through this strategic combination and division of labor, multi-agent systems have demonstrated the ability to simulate—and in some cases surpass—the collaborative efficiency of expert human teams in domains such as software development (e.g., ChatDev(Qian et al., 2024)) and scientific research (e.g., ChemCrow(M. Bran et al., 2024)).

Early explorations of GeoAI were predominantly conducted in the form of single-agent systems, validating the fundamental feasibility of leveraging LLMs for GIS tasks. For example, Li and Ning proposed the "Autonomous GIS" framework(Z. Li & Ning, 2023), in their recent study(Z. Li et al., 2025), emphasized its potential to enhance both the efficiency and innovation capacity of geospatial data processing. Zhang et al., (2024) developed Geo-GPT, which integrates

LLMs into GIS platforms， enabling the execution of predefined GIS operations through natural language instructions. Similarly, Pierdicca et al., (2025) introduced an LLM-based approach to GIS automation that transforms natural language GIS queries into executable geoprocessing workflows. By incorporating a user-friendly interface, their method effectively addresses challenges related to automation and usability in GIS workflows. C. Wei et al., (2025) further fine-tuned an LLM using a comprehensive set of pre-training data. These studies successfully demonstrated that a single LLM, acting as the "brain," can understand geospatial commands and execute corresponding tasks. Subsequently, GeoAI research has gradually advanced toward more sophisticated stages involving multi-agent collaboration. Recent works, such as ShapefileGPT by Lin et al., (2024), CartoAgent by C. Wang et al., (2025) and GIS copilot by Akinboyewa et al., (2025), have adopted multi-agent architectures involving agents with distinct roles—such as planning, execution, and evaluation—to address more complex geospatial tasks. In addition, Zhang et al.,(2025) developed a GeoAI benchmark dataset to assess the performance of existing GIS automation models. These studies mark a transition in autonomous GIS development from single-agent systems to a new era of multi-agent collaboration.

Building upon the above analysis, this study proposes GeoJSON Agents, which adopt a Planner–Worker architecture—a genuinely multi-agent framework rather than a single-agent structure such as a workflow, chain of thought (CoT), or tree of thought (ToT). Workflows, CoT, and ToT are essentially reasoning strategies and prompt-engineering techniques within a single agent. They enhance an individual LLM's problem-solving capability through structured reasoning steps or logical pathways; however, all reasoning and execution still occur within the same model and a single context. In contrast, a multi-agent architecture decomposes complex tasks into distinct functional modules that are executed by independent agents with specific roles and responsibilities. Each agent operates within its own context, equipped with tailored tools and execution abilities. These agents collaborate through well-defined communication protocols to jointly accomplish diverse user tasks.

### *2.2. Spatial Data Formats*

In geographic information science, spatial data formats function not only as carriers for data storage and exchange but also as key elements that shape the interaction between data and analytical tools. With the rise of GeoAI—and particularly the introduction of LLMs as core driving methods—the importance of this interaction paradigm has become increasingly evident. The performance of LLMs heavily depends on their ability to understand and process input data, yet the limitations of traditional formats—in terms of openness, scalability, and adaptability to task execution—are becoming more apparent. As a result, the choice of input data format has become one of the critical factors influencing the design and efficiency of GeoAI systems.

Among existing mainstream vector data formats, Shapefile is the earliest and most widely adopted. It uses a multi-file structure—typically including .shp, .shx, and .dbf files—and supports the storage of geometric types such as points, lines, and polygons, playing a foundational role in the development of GIS. However, Shapefile is inherently a binary format, which lacks human readability. It also suffers from limited field naming conventions, outdated encoding schemes, weak topological expressiveness, and high demands on storage space. Furthermore, it is incompatible with modern web environments and difficult to interface directly with AI models. GeoPackage, a newer standard format proposed by the Open Geospatial Consortium (OGC), is built on the SQLite database engine. It supports the integration of vector, raster, attribute, and metadata within a single file, offering high data integrity and strong spatial querying capabilities. However, it also uses a binary encapsulation format, which complicates direct parsing and interaction by LLMs. In contrast, GeoJSON is a lightweight format based on JSON. It has a clear structure, is easy to parse, and organizes data in a single file. It is widely used in modern WebGIS applications and embraced by the open-source community. GeoJSON encapsulates geographic features, geometric objects, and attributes as standardized JSON objects, offering excellent readability and interactivity. This aligns well with LLMs' strengths in natural language understanding and structured text processing. Additionally, GeoJSON is widely supported by major web mapping libraries such as Leaflet, Mapbox, and Kepler.gl, making it an ideal data format for modeling and executing intelligent geospatial tasks.

Based on the above analysis, the GeoJSON Agents developed in this study adopt GeoJSON as the spatial data format to fully leverage the natural compatibility of LLMs with structured text. This choice overcomes the limitations of traditional data formats and supports more automated and intelligent processing of geospatial tasks.

### *2.3.Agent Implementation Approaches*

Although the GeoJSON data format is naturally compatible with LLMs, general-purpose LLMs lack the inherent ability to directly interact with external information or domain-specific tools, making them unable to accurately perform specialized data processing and multi-step complex tasks(Brown et al., 2020; X. Huang et al., 2025; OpenAI, 2023b; Anthropic, 2024). To overcome this limitation, researchers have proposed LLM-driven agent frameworks, which enhance LLMs' ability to interact with external environments and tools through autonomous reasoning, planning, and action. Currently, the two dominant approaches to implementing intelligent agents are function calling and code generation. These methods respectively equip LLMs with greater control over execution and more flexible expressive capabilities(Mialon et al., 2023; B. Xu et al., 2023; Google. 2023).

The function calling approach, also known as tool augmentation, enables efficient interaction between agents and external tools via predefined API interfaces. Specifically, after parsing the user's natural language instruction, the agent identifies the corresponding function to be called, structures the parameters (e.g., in JSON format), and invokes external systems to execute the task and return results(Cai et al., 2024; OpenAI et al., 2024; Patil et al., 2023). Its main advantages lie in its high degree of structure and execution reliability. Because it relies on a well-defined function library, it ensures predictable task execution and consistent results. In the geospatial domain, ShapefileGPT, proposed by Lin et al.(Lin et al., 2024), is based on this method, while Krechetova & Kochedykov, (2025) constructed GeoBenchX to demonstrate the potential of LLMs equipped with function-calling capabilities for geospatial tasks. These studies collectively highlight the significant advantages of function calling in handling complex spatial analysis tasks. This approach has greatly expanded the breadth and depth of LLM applications, significantly enhancing their capability and reliability. However, the limitations of function calling are also evident. Its strong dependence on predefined function libraries constrains its flexibility in handling non-standard or highly complex tasks.

Unlike the function calling approach, the code generation method enables agents to directly translate natural language instructions into executable code. By dynamically generating program code (e.g., Python), the agent can complete user-defined tasks(M. Chen et al., 2021; Nijkamp, Hayashi, et al., 2023; Rozière et al., 2024; Google. 2024). Specifically, this method involves semantic understanding and decomposition of user instructions, generation and execution of dynamic code, and self-correction through feedback and iterative refinement — forming a continuous and adaptive closed-loop process(X. Chen et al., 2023; Shinn et al., 2023). The core strengths of the code generation method lie in its exceptional flexibility and generalizability, making it particularly suitable for tasks requiring innovative algorithm design or non-standard, complex problem solving. For instance, Hou et al., (2025) proposed GeoCode-GPT, a pre-trained model dedicated to geospatial code generation. They subsequently introduced GeoCogent(Hou, Jiao, et al., 2025) , the first LLM-based intelligent framework for geospatial code generation, which integrates planning, tool utilization, and memory mechanisms to enable efficient code generation for geospatial tasks. These applications collectively demonstrate the great potential of code generation methods in advancing geospatial analysis. However, despite its vast potential, the code generation approach faces challenges such as high execution uncertainty and difficulties in ensuring system stability and security. The generated code may contain semantic or logical flaws, or introduce security vulnerabilities.

Although both function calling and code generation have made significant progress, their respective performance in geospatial analysis tasks—particularly in handling complex tasks based on the GeoJSON data format—remains insufficiently studied in a systematic manner. To address this research gap, this study constructs two LLM-driven multi-agent models—GeoJSON Agents—based on function calling and code generation, respectively. It also introduces a GeoJSON task benchmark covering multiple levels of task complexity, aiming to systematically evaluate the performance of both methods in geospatial analysis and to provide a solid theoretical foundation and empirical support for the development of autonomous GIS systems.

## 3. Methodology

### *3.1. Overview*

In the field of artificial intelligence, agent is an autonomous system capable of perceiving its environment, making decisions, and taking actions to achieve specific objectives. Building upon this concept, a multi-agent system comprises multiple specialized agents that collaborate through division of labor to tackle complex tasks, often achieving better performance than a single agent. To address the dual limitations of traditional GIS operations and general-purpose LLMs in geospatial analysis, this study designs and implements an automated spatial analysis framework—GeoJSON Agents. Centered around the modern and lightweight GeoJSON data format, the framework enables the execution of complex spatial tasks via natural language instructions. It adopts an LLM-driven multi-agent architecture that separates task planning from execution, thereby enhancing system reliability and flexibility. Specifically, the architecture of GeoJSON Agents consists of two core agents—the Planner and the Worker—both powered by LLMs and interconnected through a continuous feedback loop, forming an LLM-driven closed system. The Planner agent focuses on reasoning and task decomposition, while the Worker agent is responsible for tool utilization through function calls or code generation. This LLM-supported feedback mechanism enables iterative optimization of geospatial analysis execution through continuous interaction between the planning and execution phases.

At the implementation level, this study employs two distinct LLM-enhancement strategies—function calling and code generation—to construct the Worker agent, and further explores the performance of each approach. The function-calling-based GeoJSON Agent leverages a closed library of 40 APIs, ensuring reliability and stability throughout the execution process. In contrast, the code-generation-based GeoJSON Agent enables the Worker agent to dynamically generate Python code and incorporates a closed-loop self-correction mechanism, thereby enhancing the system's flexibility and generalizability when dealing with open-ended and complex tasks. To objectively evaluate and compare the performance of these two implementation methods, a hierarchical benchmark was constructed, comprising 70 task cases across three levels of complexity: basic, intermediate, and advanced. The benchmark tasks are sourced from authoritative GIS textbooks and cutting-edge research to ensure the validity and representativeness of the evaluation. Through systematic experimental evaluation, the study confirms the effectiveness and applicability of both methods in geospatial task processing, providing strong support for advancing the automation and intelligence of geospatial analysis.

### *3.2. Hierarchical Task Benchmark Design*

#### *3.2.1. Task Case Design*

To objectively and comprehensively evaluate the performance of GeoJSON Agents in spatial analysis tasks, this study constructs a hierarchical geospatial task benchmark comprising 70 task cases spanning various levels of complexity (see Appendix3-5). The benchmark is designed along two orthogonal dimensions: the breadth of task types and the depth of task complexity, resulting in a relatively comprehensive dataset.

Along the task type dimension, to ensure both coverage and representativeness of the benchmark, we synthesized 40 core operational tasks in geographic information systems (GIS), drawing on relevant literature(Sui, 2009; Guoan Tang, 2017). Following the classic classification of spatial analysis tasks in(Sui, 2009), these 40 operations were further categorized into six major groups: spatial data acquisition, spatial data storage and transformation, spatial data interaction, spatial relationship analysis and applications, spatial query and retrieval, and spatial cartography (see Table 1 and Appendix1).

**Table 1. Number of tasks per spatial analysis category**

| Task Category | Number |
|---|---|
| Spatial Data Acquisition | 2 |
| Spatial Data Storage and Conversion | 5 |
| Spatial Data Interaction | 2 |

| Task Category | Number |
|---|---|
| Spatial Relation Analysis And Application | 25 |
| Spatial Query And Retrieval | 5 |
| Spatial Mapping | 1 |
| Total | 40 |

Along the task complexity dimension, to systematically evaluate the capabilities of the two GeoJSON Agents, the 70 task cases were categorized into three difficulty levels based on their internal logic, number of procedural steps, and the degree of autonomous planning required from the model:

**Basic tasks**: This category includes 40 cases designed to simulate atomic, single-step spatial analysis operations. Each case typically breaks down into only one or two subtasks, and is intended to assess the model's ability to precisely interpret and execute fundamental geospatial tasks. Representative examples include: coordinate system transformation, calculating geometric lengths, querying by a single attribute, and generating simple point, line, or polygon features.

**Intermediate tasks**: This level includes 20 cases requiring the model to understand and execute multi-step linear workflows. Success depends not only on the accuracy of individual steps but also on the model's ability to decompose tasks, maintain contextual state, and chain together various tools. A representative example is: "First, filter counties in Pennsylvania where household income exceeds $100,000, then use the filtered results to create a bar chart of household incomes in those counties."

**Advanced tasks**: This category includes 10 cases that simulate realistic, open-ended analytical scenarios, placing greater demands on the model's autonomous planning and complex problem-solving abilities. Characterized by ambiguous instructions and compound objectives, these tasks require the model to autonomously construct a complete analytical workflow that integrates task decomposition, logical reasoning, and evaluation. Representative examples include multi-criteria site suitability analysis or thematic regional mapping and analysis (e.g., building density).

### *3.2.2. Data Format and Dataset*

At the data format level, this study uniformly and exclusively adopts GeoJSON as the sole format for all task cases. This choice is motivated by the high compatibility between the GeoJSON format and LLM-driven automated workflows (see Section 2.2 for a detailed discussion on spatial data formats).

In selecting datasets, we adhered to two key principles: authoritativeness and representativeness of real-world scenarios. The datasets employed in this study are derived from two authoritative sources: (1) the pioneering GeoAI research paper GIS Copilot: Towards an Autonomous GIS Agent for Spatial Analysis(Akinboyewa et al., 2025), and (2) the experimental data accompanying the renowned GIS textbook authored by Professor Guoan Tang and collaborators (Guoan Tang, 2017). These datasets include a diverse range of feature types—points, lines, and polygons—and span various spatial scales, from points of interest (POIs) to road networks. As all datasets were originally in Shapefile format, we performed lossless format conversion to transform them into GeoJSON, ensuring compatibility with the framework.

This rigorous control over the authoritativeness and representativeness of data sources fundamentally ensures the authenticity and validity of the benchmark tasks, providing a solid and reliable foundation for subsequent model performance evaluation.

### *3.3. Model Architecture and Implementation*

#### *3.3.1. Overall Design: A Multi-Agent Architecture for GeoJSON Agents*

To effectively address the inherent complexity and multi-step nature of geospatial tasks, this study designs and implements a unified multi-agent architecture—GeoJSON Agent (see Figure 1)—that integrates both function-calling and code-generation approaches. The architecture consists of two core agents: the Planner and the Worker. It is designed to overcome the limitations of single-agent structures such as workflows, chain of thought (CoT), and tree of thought (ToT), in which all reasoning and execution are performed by the same model within a single context. Such single-agent systems often suffer from "hallucinations" when handling lengthy reasoning processes, making them less effective for complex, multi-step geospatial tasks(Masterman et al., 2024). In contrast, the Planner–Worker architecture of the GeoJSON Agent establishes a clear division of labor and collaboration mechanism. The Planner agent focuses on high-level task understanding, decomposition, and dynamic planning. It does not directly manipulate data or invoke tools; instead, it translates ambiguous user instructions into well-defined sequences of subtasks and communicates them to the Worker. The Worker agent, in turn, concentrates on low-level operations such as function calls or code generation. It performs concrete geospatial operations within an isolated execution environment and returns the results to the Planner. This division of responsibilities and cooperative mechanism offers significant advantages: (1) Decoupled roles – separating planning and execution mitigates the "hallucination" issue common in single-agent systems with long reasoning chains. The Planner can focus on understanding user intent, performing logical reasoning, and decomposing tasks, while the Worker can accurately execute subtasks in an independent environment and report the outcomes. (2) Dynamic collaboration – the Planner can adaptively adjust subsequent subtasks and plans based on real-time feedback from the Worker, forming a closed-loop, self-adaptive execution process. This is particularly crucial for managing highly uncertain and complex tasks. Through this structured division and iterative coordination, the GeoJSON Agent architecture achieves both the flexibility of high-level planning and the reliability of low-level execution, enabling efficient handling of geospatial analysis tasks ranging from simple geometric operations to complex multi-step analyses.

**Planner**: The Planner agent is equipped with strong perception, reasoning, and memory capabilities. It accurately interprets user intent and intelligently decomposes complex instructions into detailed subtasks. It employs reasoning strategies such as Chain-of-Thought (CoT) ((J. Wei et al., n.d.), Tree-of-Thought (ToT)(Yao et al., 2023), and Graph-of-Thought (GoT)(Besta et al., 2024)to perform effective task decomposition and dynamic adjustment(C. Wang et al., 2025). Within the GeoJSON Agent architecture, the Planner is responsible for the following:(1) Intent Interpretation and Task Decomposition: Accurately parses the user's natural language instructions and decomposes complex tasks into a logically ordered sequence of simpler subtasks. (2) Execution Monitoring and Dynamic Adjustment: Continuously receives and evaluates feedback from the Worker during task execution, dynamically updating subsequent plans and subtasks until the overall task is completed and the final result is returned to the user.

**Worker**: The Worker is responsible for executing the assigned subtasks. After selecting and invoking the appropriate tools, it performs operations within an isolated sandbox environment, ensuring safe execution without interfering with other processes or compromising data security. It completes each task carefully and provides real-time feedback. Its core responsibilities include: (1) Instruction Reception and Execution: Receives and executes clear subtask instructions from the Planner. (2) Tool Invocation: Calls authorized tools according to the given instructions and interacts with the external environment. In this study, the notion of "tool" is a critical variable: in the function-calling-based model, it refers to a predefined function library (API); in the code-generation-based model, it refers to a Python code interpreter capable of executing dynamically generated scripts. (3) Status and Result Feedback: Upon completing each task, the Worker promptly and accurately sends the execution outcome—whether successful results or error messages—back to the Planner in real time.

The two agents collaborate through a closed-loop feedback system, following a cyclical process of "planning → execution → observation → re-planning." Specifically, whenever a user uploads task instructions and input files, the Planner first interprets and decomposes potentially ambiguous natural language instructions into multiple subtasks, then assigns these subtasks to the Worker. Upon receiving the Planner's instructions, the Worker selects and applies the most appropriate tools for execution, and subsequently returns the results to the Planner. Based on this feedback, the Planner updates its understanding of the task status and formulates the next optimal subtask for the Worker, continuing the cycle until all subtasks are completed. This model, which integrates dynamic planning with feedback loops, not only provides greater robustness in handling long and complex tasks but also significantly enhances the interpretability and controllability of the entire system(Kim et al., 2024). To provide a more intuitive illustration of the architecture's operational process, Figures 9–11 (see Section 4.3, Case Studies) present the complete workflows of the function-calling-based and code-generation-based GeoJSON Agent across tasks of varying complexity—basic, intermediate, and advanced. These workflows encompass the Planner's task decomposition strategies, the Worker's specific execution steps, and the interaction patterns between the two agents. Collectively, these cases clearly demonstrate how the multi-agent architecture achieves effective collaboration to accomplish a wide range of geospatial tasks, from simple buffer analyses to complex thematic map generation.

### *3.3.2. Implementation Details*

To elucidate the underlying operational mechanism of the GeoJSON Agents framework, this section explains the system's implementation from a methodological perspective and describes it from three aspects: the model's operational mechanism and workflow, the prompt-based interactions between agents, and the integration of LLMs through API calls.

**Runtime Mechanism and Execution Flow**: The GeoJSON Agents framework is a Python-based LLM system that interacts with the model through OpenAI's ChatCompletion API (in this study, gpt-4o-2024-11-20). Both the Planner and Worker agents are powered by LLMs. When a user submits a task request along with a GeoJSON file, the system initializes an execution environment to maintain task states, conversation history, and intermediate results. The Planner first retrieves file information and relevant data, and based on this contextual information, decomposes the user's request into a sequence of subtasks. Each subtask is then transmitted to the Worker as an independent prompt instruction. Upon receiving a subtask, the Worker executes the corresponding operation according to the implementation method. In the code-generation-based GeoJSON Agent, the Worker employs the execute_code tool to generate and run Python scripts. If execution fails, the Worker autonomously analyzes the error messages, revises the code, and re-executes it—up to five attempts. In the function-calling-based GeoJSON Agent, the Worker selects appropriate functions from the API documentation embedded within the prompt, formulates standardized function calls, and the system parses and executes the corresponding predefined functions. After execution, the Worker returns the results to the Planner. The Planner evaluates the output and determines the next course of action: if the previous subtask succeeds, it issues the next one; if it fails, it instructs the Worker to retry. Once all subtasks are completed, the Planner sends an "END" termination signal to conclude the process. This plan–execute–evaluate loop continues iteratively until the entire task is successfully completed.

**Agent Interaction Through Prompts**: All interactions between the Planner and Worker are conducted through carefully designed prompts. These prompts not only define the roles and responsibilities of each agent but also specify the available tools, behavioral guidelines, and output formats. They serve as the sole communication channel between the agents, ensuring consistent coordination and controlled task execution.

The Planner's prompt defines its role as a "professional AI assistant with 20 years of experience in GIS," and clearly specifies its core responsibilities: task decomposition, instruction generation, and result evaluation (see Appendix 6.1 for the full prompt). The prompt also delineates the available tools and their appropriate usage contexts, while providing explicit guidelines for task decomposition—emphasizing the avoidance of over-fragmentation and the maintenance of logical independence among subtasks. The overall execution process is structured into seven stages, ranging from initial task analysis to the issuance of the final "END" message.

The design of the Worker's prompt differs significantly between the two implementation methods. In the function-calling-based *GeoJSON Agent*, the Worker's prompt embeds a complete YAML-formatted documentation of 40 API functions (see Appendix 6.3 for the full prompt), which serves as the sole reference for function selection. The prompt enforces nine strict constraints, including: only the listed APIs may be used; each step must return a single function call; function calls must follow a prescribed format; the first step must read the data; and the final step must save the results, *etc.* These constraints ensure the standardization and predictability of function calls. In contrast, the code-generation-based *GeoJSON Agent* emphasizes the Worker's autonomy and accountability. Its prompt explicitly states that the Worker is "fully responsible for ensuring successful code execution" and must "independently debug within the specified number of attempts until the task is successfully completed," without delegating execution to any other agent (see Appendix 6.2 for the full prompt). This design grants the Worker a high degree of autonomy, enabling iterative code refinement within the allowed number of attempts without external intervention. The prompt also includes detailed error-handling strategies—for example, when encountering file path errors, the Worker is required to use available tools to locate the correct path.

A key feature of the interaction mechanism between the Planner and the Worker is that they never communicate directly. When the Planner needs to send a subtask to the Worker, the system extracts the subtask description generated by the Planner, combines it with the Worker's system prompt to construct a new dialogue context, and transmits it through a separate LLM API call. After the Worker's response is executed by the system, the resulting output is combined with the Planner's prompt and returned via another API call. This design transforms inter-agent communication into a series of independent LLM reasoning processes, each carried out within a carefully constructed prompt context.

**API-Based LLM Integration:** All intelligent behaviors within GeoJSON Agents—including task planning, code generation, and function selection—are realized through API calls to the LLM. The system adopts a model-agnostic architecture; however, this study primarily employs OpenAI's gpt-4o model for experimental validation. How the Planner decomposes tasks, how the Worker selects functions or generates code, and how both respond to errors depend entirely on the LLM's reasoning ability based on prompt context. The system implements a unified LLM client module to encapsulate and manage API interactions between agents. Each communication between the Planner and Worker is represented as two independent API calls: the Planner's output is parsed and reformulated as the Worker's input prompt for transmission, and the Worker's response is similarly processed and returned as the Planner's input prompt. This mechanism enables bidirectional information exchange and forms a logical closed loop between the two agents. In the function-calling approach, the Worker returns structured function-call outputs via the API. The system extracts the function names and parameters, verifies them, and executes the corresponding operations within an isolated environment. The execution results are then appended to the conversation history, forming a complete "call–execute–feedback" cycle. In the code-generation approach, the Worker outputs Python code, which is executed in a sandbox environment preconfigured with geospatial libraries. If execution fails, the system captures detailed error information and triggers a new API call to prompt the Worker to revise the code—iterating up to five times until successful completion. This API-based interaction architecture provides the model with high flexibility and adaptability to complex tasks. To enhance traceability and continuous optimization, the system includes a comprehensive logging mechanism that records the content of each API prompt, model response, response time, function-call parameters, generated code files, and execution logs.

### *3.3.3. Function-Calling-Based GeoJSON Agent*

In the function-calling-based GeoJSON Agent, the model architecture is formalized as a highly structured execution system centered around a predefined function library. The primary objective of the GeoJSON Agent is to translate user-provided natural language instructions into a reliable and precise sequence of API calls from the closed function library. Within the model, the Planner is explicitly positioned as a task planner responsible for decomposing complex user instructions into a logically ordered set of structured subtasks. The core innovation of the model lies in the design of the Worker, which functions not as a general executor but as a highly specialized API execution agent. All of its operations take place within an execution environment composed of three key components, ensuring both accuracy and reliability in task execution: 1) a predefined function library; (2) an associated API documentation file specifying parameter rules and formats; and (3) metadata of the files to be processed, used to prevent operational errors. Upon receiving instructions from the Planner, the Worker leverages these environment components to select the most appropriate function, generate a function call with precise parameters, and safely execute it within an isolated sandbox environment.

**Function Library and API Documentation**: The core mechanism of the proposed model is the function calling approach. Function calling is a technique designed to enhance the interaction between LLMs and external programs, allowing the model to invoke predefined and deterministic functions while generating text. Compared to the more open-ended code generation approach, function calling offers a closed yet reliable set of tools, providing a more controlled solution pathway for handling complex tasks.

Given the advantages of the function calling method in terms of controllability and stability, this study adopts it as one of the core implementation strategies for the GeoJSON Agent. The goal is to enhance execution reliability and domain specificity when handling structured geospatial data and performing specialized spatial operations, thereby improving task completion accuracy. Through a carefully designed and customized function library, the function calling method effectively constrains and guides LLM behavior. By operating within a defined scope, it encourages accurate function matching and invocation, offering a more dependable solution for geospatial tasks. Moreover, this approach significantly reduces the risk of task failure due to semantic misinterpretation or logical errors—common in general-purpose LLMs lacking domain expertise. In geospatial data processing scenarios that demand high consistency and procedural rigor, function calling demonstrates superior controllability and operational safety.

To achieve this goal, we developed a function library specifically tailored to the GeoJSON data structure for the model's Worker component. In designing the function library, we adhered to the principles of systematicity, authoritativeness, and comprehensiveness to ensure that it provides a necessary and sufficient set of tools for completing the benchmark tasks. Specifically, the selection and design of the 40 functions were not based on subjective choice but on a systematic review and synthesis of authoritative GIS textbooks (Guoan Tang, 2017). and mainstream GIS software toolboxes, such as QGIS. We first identified six major categories of core spatial analysis operations from the textbook—spatial data acquisition, spatial data storage and transformation, spatial data interaction, spatial relationship analysis and application, spatial query and retrieval, and spatial cartography (see Table 1). We then systematically mapped and integrated the corresponding functionalities from the QGIS toolbox to these categories, ultimately forming a standardized set of 40 fundamental and intermediate GIS operations. This library includes 40 function APIs with specialized geospatial capabilities (see Appendix 2), accompanied by comprehensive API documentation. The documentation provides detailed descriptions for each API, including function names, parameter definitions and types, usage instructions, and practical examples. This offers clear guidance on tool functionality and assists the Worker in accurately matching and invoking the correct functions. In addition, to balance real-time LLM guidance with rigorous system-level verification, the API documentation is stored in both YAML and JSON formats (see Figure 2). The YAML format, due to its conciseness and human readability, is embedded into the LLM's context to guide real-time function invocation. The JSON format, with its clear key-value structure, serves as an ideal medium for system-level interactions and validating the correctness of function calls.

The function library is designed to equip large language models with the essential capabilities required for executing GIS-related geospatial tasks. Its functionality spans a wide range—from basic data transformation and processing (e.g., renaming fields, converting file formats) to critical spatial relationship analyses such as spatial joins, buffer generation, and creation of Thiessen (Voronoi) polygons. During task execution, the Worker selects the most appropriate API from the function library based on instructions from the Planner and real-time execution feedback. For example, in handling spatial query tasks, the model first evaluates the necessary lookup steps, then sequentially matches and invokes the corresponding APIs as the task progresses. This approach not only avoids redundant operations but also emulates the structured, step-by-step workflow followed by GIS professionals, thereby ensuring the accuracy of task outcomes.

```yaml
Function Name:
  usage: {usage}
  type: {type}
  param:
  -param A:
    type: {type of param A}
    description: {description of param A}
  -param B:
    type: {type of param B}
    description: {description of param B}
 Example:
  # Example 1:
     description: {description of task A}
     calling: {function A}
  # Example 2:
     description: {description of task B}
     calling: {function A}
```

YAML

```json
{
  "{Function Name}":{
     "usage":"{usage}",
     "type":"{type}",
     "parameters":{
       "param A":{
       "type":"{type of param A}",
       "description":"{description of param A}"
     },
       "param B":{
       "type":"{type of param B}",
       "description":"{description of param B}"
     }
   },
 "Example":[
   {
     "# Example 1":
       "description":"{description of task A}"
       "calling":"{function A}"
   },
   {
     "# Example 2":
       "description":"{description of task B}"
       "calling":"{function A}"
   }
  ]
 }
}
```

JSON

**Figure 2. The function library documentation in YAML (left) and JSON (right) formats.** This figure illustrates the dual-format design of the function library documentation used in GeoJSON Agents. The YAML schema (left) provides a concise and human-readable representation for in-context learning of large language models, specifying the function name, usage, type, parameters, and example calls. In contrast, the JSON schema (right) translates the same information into a machine-readable hierarchical structure suitable for external system interaction, function invocation, and automated result validation. Together, these two representations ensure both interpretability for LLM-based reasoning and interoperability with downstream geospatial processing components.

**Workflow:** The function-calling-based GeoJSON Agent follows a highly structured planning and execution loop centered on a predefined function library(see Figure 3). Through the seamless integration of the Planner and Worker, the system forms a complete and autonomous task-handling pipeline. Upon receiving the user's task instruction and the uploaded GeoJSON file, the system immediately initiates the Planner's planning loop and initializes a working environment to record task progress, current status, and historical information. In each planning cycle, the Planner first evaluates the current task state to determine whether the user's overall objective has been met. If not, it initiates the subtask generation process, breaking down the high-level objective into multiple subtasks that can be mapped to specific APIs in the function library, and assigns them sequentially to the Worker. Upon receiving a subtask, the Worker operates within a defined execution environment comprising the function library (for selecting tools), the API documentation (for parameter configuration), and the metadata of the input file (for error prevention). Based on this information, the Worker selects the most appropriate API for the given subtask, generates a function call with precise parameters, and safely executes it within an isolated sandbox environment. After execution, the Worker returns either the result or any encountered error information to the Planner. The Planner then updates its working environment based on the feedback and decides whether to proceed with the next subtask in the plan or replan based on the reported error. This forms a dynamic, closed-loop, and iterative workflow that continues until the task is completed and the final result is returned to the user.

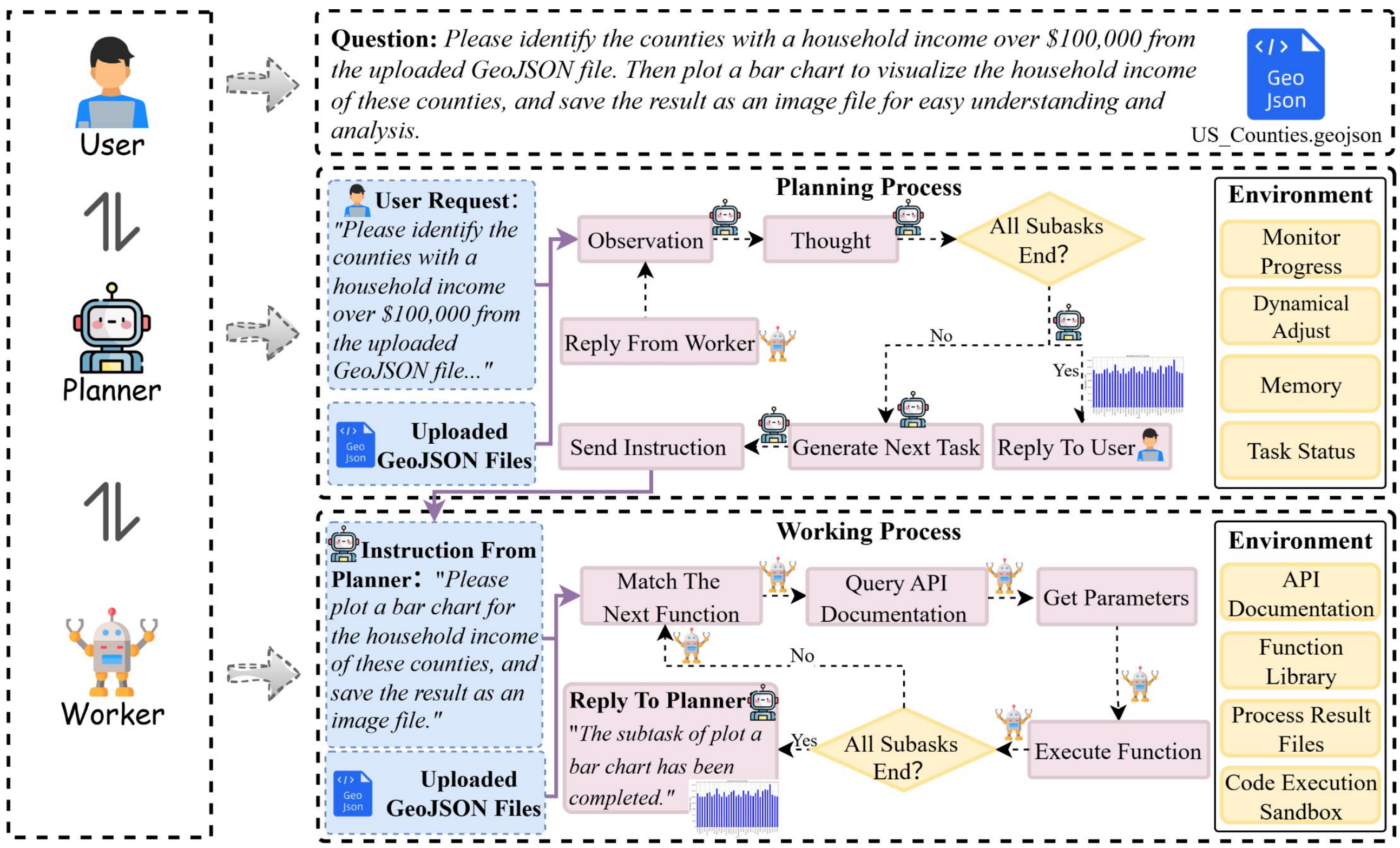

**Figure 3. Workflow of the function-calling-based GeoJSON Agent.** User uploads spatial task instructions and input files. Planner converts the user's ambiguous instructions into a set of decomposed subtasks and assigns them to the Worker. For each assigned subtask, Worker selects the most appropriate API, generates a function call instruction with precise parameters, and executes it. then returns the results to the Planner.

### *3.3.4. Code-Generation-Based GeoJSON Agent*

The code-generation-based GeoJSON Agent implements a dynamic and flexible architecture that eliminates the dependence on closed, predefined function libraries. Instead, it enables the Worker to dynamically generate, execute, and debug Python scripts to accomplish the subtasks planned by the Planner. Within this architecture, the Planner serves as the "brain" of the system—analyzing and deconstructing user requirements and available data into a sequence of logically coherent subtask instructions, while dynamically adjusting the workflow as needed. Correspondingly, the Worker functions as the system's "hands." At its core lies a code generator capable of producing Python scripts for specific subtasks using specialized libraries such as GeoPandas and Shapely, which are then executed within an execute_code sandbox environment. To further enhance system performance, a self-repair and debugging mechanism is integrated within the model. When code execution fails, the Worker's built-in debugging module autonomously analyzes the returned error information, identifies the source of the issue, and iteratively revises and re-executes the code (with a maximum of five retries in this study). In addition, a structured logging system records every execution cycle and generated script, ensuring full traceability, process transparency, and reproducibility. Collectively, these tightly integrated modular components form a powerful, flexible, and robust working mechanism that enables the model to effectively handle a wide range of complex geospatial analysis tasks.

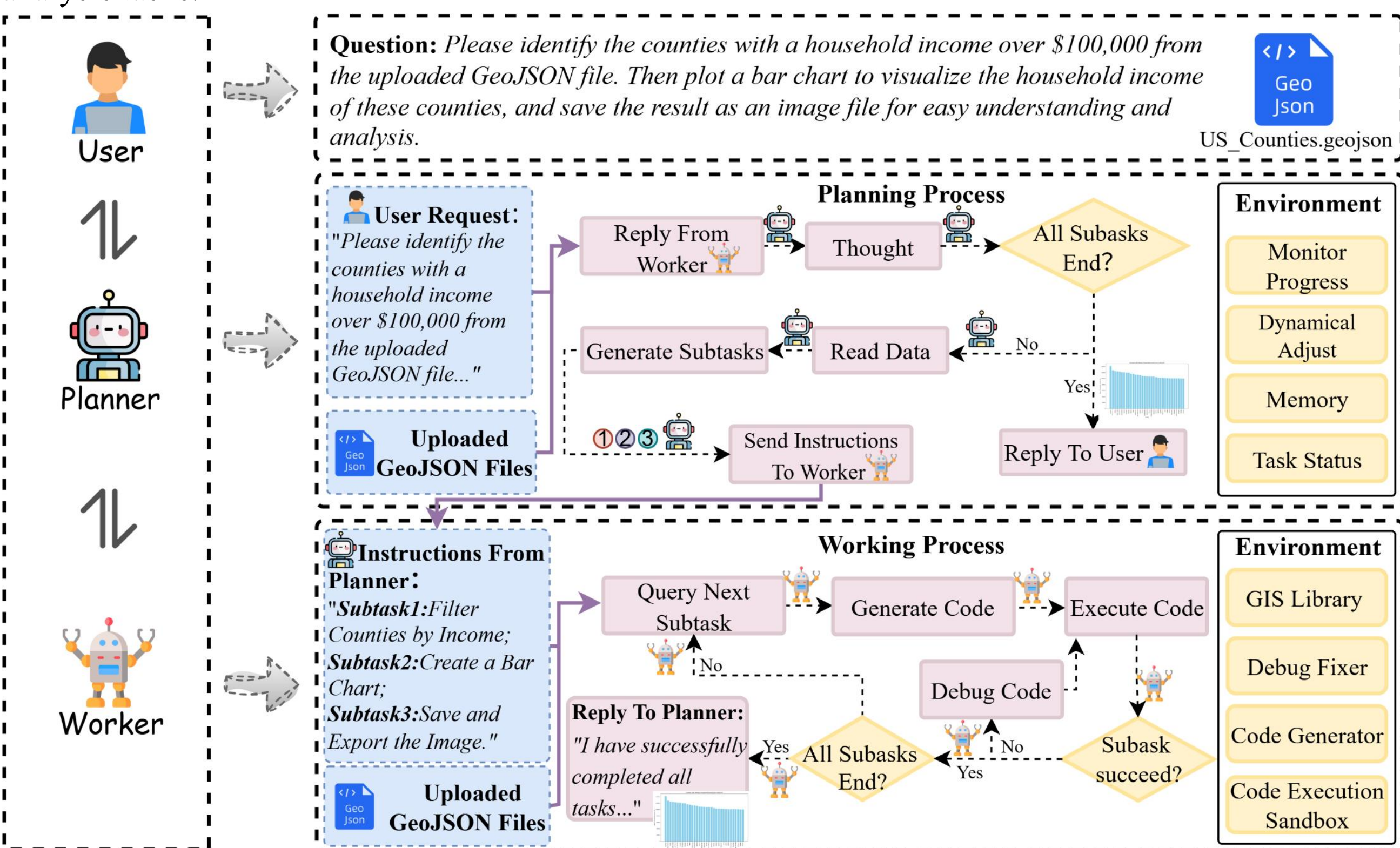

**Figure 4. Workflow of the code-generation-based GeoJSON Agent.** User uploads spatial task instructions and input files. Planner converts the user's ambiguous instructions into a set of decomposed subtasks and assigns them to the Worker. For each assigned subtask, Worker generates the corresponding Python code and executes it using the execute_code tool, then returns the results to the Planner.

**Dynamic Code Generation and Closed-Loop Debugging:** The core of the code generation model lies in its dynamic code generation capability and an innovative closed-loop self-repair and debugging mechanism, which together enable the model to "create solutions in real time."

**(1) Dynamic Code Generation**

In the code generation-based GeoJSON Agent, once the Worker receives a subtask instruction from the Planner, the embedded LLM assumes the role of a "GIS-savvy programmer" and dynamically generates an executable Python script tailored to that subtask. To ensure domain relevance and efficiency, the system prompts the LLM to prioritize widely recognized geospatial libraries such as GeoPandas (for vector data operations), Shapely (for geometric computations), and Matplotlib (for data visualization). This process enables the generation of customized, one-time-use scripts for each subtask, offering a high degree of flexibility.

**(2) Closed-Loop Self-Repair and Debugging**

To handle inevitable errors in dynamically generated code, the Worker is equipped with a closed-loop self-repair and debugging mechanism, following an iterative "execute–validate–repair" cycle (see Figure 4). Step 1: Execution and Error Capture – The generated code is executed in a secure execute_code sandbox. If the execution fails, the system does not terminate the task. Instead, it captures the complete error traceback returned by the Python interpreter (e.g., SyntaxError, KeyError, or FileNotFoundError). Step 2: Self-reflection and correction – The Worker takes the captured error message, the original failed code, and the initial subtask instruction as input and feeds them back into itself. At this stage, the Worker acts as a "code debugging expert," analyzing the root cause of the error and, based on this analysis, modifying and regenerating a new version of the code. Step 3: Iterative Retry and Termination – The updated code is re-executed in the sandbox. This cycle continues to address successive errors. To avoid infinite loops, the system limits retries to five attempts. If all attempts fail, the Worker reports the subtask as failed to the Planner. This closed-loop repair mechanism allows the model to autonomously diagnose and correct errors, effectively mitigating subtle flaws in the code generation process and significantly improving the system's success rate and overall robustness in complex task execution.

**Workflow**: The workflow of the code generation-based GeoJSON Agent begins when a user submits a task request along with the corresponding GeoJSON data files. Next, the Planner invokes a file analysis module to inspect the available data and infer its structure. Based on this analysis and the user's instruction, the Planner systematically decomposes the complex task into a sequence of logically coherent subtasks. Each subtask is then sequentially dispatched to the Worker. The Worker generates a customized Python script tailored to the specific subtask and executes it within a secure execute_code sandbox. If execution fails, the Worker analyzes the underlying cause and iteratively refines the code until successful execution is achieved. Upon success, the result and execution status are reported back to the Planner. The Planner evaluates the returned output; if it aligns with the expected outcome, the next subtask is scheduled. Otherwise, the Planner requests the Worker to revise and re-execute the current step. This plan–execute–evaluate loop continues iteratively. After completing each subtask, the Worker stores the intermediate results, which are later used by downstream subtasks, ensuring smooth and context-aware task progression. Once the full task chain has been successfully executed, the Planner issues an "END" signal. The system then finalizes the process by saving the final outputs (e.g., generated maps, analytical results, or visualizations) to designated files and records the full execution log for reproducibility and audit purposes.

### *3.4.Experimental Design*

**Dataset:** This study utilizes the dataset introduced in Section 2 as part of the task benchmark. The dataset consists of data files in GeoJSON format and encompasses a range of spatial tasks, including data storage and transformation, spatial relationship analysis, spatial queries, and geometric computations. It is specifically designed to evaluate the capabilities of two GeoJSON Agents—one for function calling and the other for code generation—in handling vector data of varying types and complexities, as well as in executing the corresponding operations.

**Baseline Models:** To comprehensively evaluate the performance of the two GeoJSON Agents, we conducted comparative experiments against several state-of-the-art general-purpose large language models (LLMs). The selected baselines include GLM-4-2024-0520 (Zhipu AI), GPT-4o-mini-2024-07-18, and GPT-4o-2024-11-20 (OpenAI), as well as their respective 2025 iterations—GLM-4.6-2025-09-30 and o4-mini-high-2025-04-16. These models represent the leading capabilities of contemporary general-purpose LLMs. To ensure fairness and comparability in evaluation, we employed the official interaction interfaces and online chat systems provided by each platform, and established standardized evaluation workflows. All experiments were conducted via the official chat interfaces of the baseline models, without the use of any third-party plugins or extended APIs, thereby simulating real-world user scenarios. The evaluation process for each baseline model followed a consistent set of steps. First, we uploaded the task files containing GeoJSON data to the file system supported by each model (e.g., OpenAI's file API or GLM's file upload interface). Simultaneously, we entered a standardized spatial task prompt in natural language into the chat interface, guiding the model to interpret the task intent and generate code capable of performing the required spatial analysis. Next, the model executed the generated code within its corresponding sandbox environment, accessed the uploaded data files, and performed the designated operations. Finally, the analysis results were returned either in structured natural language or as newly generated data files, enabling further assessment and comparison. To ensure consistency and rigor in model comparison, we adopted a unified multi-attempt execution strategy for all models, including the two GeoJSON Agents and five baseline general-purpose LLMs. Specifically, each task was executed up to five times. A task was considered successful if it was completed at least once within these five attempts; conversely, it was labeled as failed only if all five attempts were unsuccessful. This design effectively balances two objectives: on the one hand, it provides models with sufficient opportunities to overcome transient errors caused by the probabilistic nature of LLM outputs; on the other hand, it prevents excessive retries that could inflate computational costs and distort evaluation results, thereby making the assessment more representative of real-world performance. During execution, external intervention was strictly controlled. No modifications, clarifications, or follow-up questions were allowed for task prompts. Each retry had to rely solely on the model's own contextual understanding and self-correction capability. This constraint ensured fairness across models in both the number of attempts and evaluation criteria, while eliminating potential bias introduced by human intervention. Otherwise， although the baseline models leveraged their default built-in code interpreters during task execution, no external plugins or custom APIs were introduced in our experiments. All operations were conducted using only the officially available interaction capabilities provided by each platform. This experimental framework was designed to simulate realistic application scenarios in which general-purpose LLMs operate within their native environments and configurations to directly perform geospatial tasks. It enables a systematic comparison between these general-purpose models and the two customized GeoJSON Agents proposed in this study across key dimensions such as spatial reasoning ability, task comprehension accuracy, execution capability, and overall stability.

**Evaluation Metrics:** Accuracy was used as the primary metric to assess the reliability and effectiveness of both the baseline models and our GeoJSON Agents in performing GeoJSON-based spatial tasks. Additionally, we employed the average number of execution rounds as a metric to evaluate the models' level of automation and operational efficiency. A task was considered successful if the model, without any human intervention, produced an output file that was logically and semantically correct and met all task requirements. A task was deemed a failure if it remained unsuccessful after five attempts. Accuracy represents the proportion of tasks successfully completed out of the total number of tasks, serving as the definitive measure of each model's ability to perform the assigned work. The average number of execution rounds measures how many complete interaction cycles the model required to successfully complete a task. An execution round is defined as a full interaction cycle, beginning when the Planner issues a subtask instruction to the Worker and ending when the Planner receives the final feedback from the Worker. For successfully completed tasks, a lower average number of execution rounds—particularly values approaching 1—indicates greater model stability and efficiency.

**Evaluation Criteria**: To ensure accuracy, objectivity, and rigor in the experimental results, this study employed a systematic validation procedure during the model evaluation stage. It should be noted that some of the test tasks were derived from the GIS Copilot study (Akinboyewa et al., 2025). For these tasks, the standard answers provided in that study were adopted as reference outputs, and our model results were compared against them to determine task correctness—further enhancing the objectivity and comparability of the evaluation. For all other tasks, validation methods were adapted to the type of task output. For tasks generating structured data files (e.g., spatial joins or field renaming), the model outputs were manually inspected to verify: (1) data structure integrity (field names, types, and counts), (2) correctness of feature numbers, (3) accuracy of attribute values (for numeric fields, whether calculated results matched expected values; for string fields, whether contents were correctly aligned), and (4) validity of geometry types and coordinate reference systems. For tasks involving spatial geometry operations (e.g., buffer analysis or Thiessen polygons), the generated geometries were loaded into QGIS for visual overlay inspection. Geometric correctness—including shape, spatial location, and topological relationships—was verified through visual assessment, while key geometric measures (e.g., buffer area, line length, etc.) were validated using QGIS measurement tools or GeoPandas geometry functions to ensure they fell within acceptable error ranges. For tasks generating charts or maps, outputs were manually examined against the source data to confirm that the visualization correctly represented the data, that chart elements (axes, labels, legends, etc.) were complete and accurate, and that the overall visual presentation met task requirements. A task was marked as successfully completed only when all relevant validation criteria were satisfied. For complex tasks that allow multiple valid implementation pathways, the evaluation focused on whether the output functionally achieved the intended geospatial analysis objectives, rather than enforcing strict consistency in format or implementation details. This classification-based evaluation framework ensures that the accuracy metric faithfully reflects each model's true capability to perform geospatial analysis tasks.

## 4. Experimental Results

In this section, we systematically present the performance of the two GeoJSON Agents. Specifically, Section 4.1 presents a comparative analysis between the two GeoJSON Agents and general-purpose large language models in executing GeoJSON-based spatial tasks, aiming to assess whether the proposed agents offer significant improvements in GIS spatial analysis. Section 4.2 examines the performance of each GeoJSON Agent across tasks of varying complexity, compares the results of the two agents, and provides an in-depth analysis of the performance differences between the two core approaches: function calling and code generation. Section 4.3 demonstrates the effectiveness of the two GeoJSON Agents through case studies involving tasks of different complexity levels. Section 4.4 presents comparative experiments involving GeoJSON Agents configured with different LLMs. Finally, Section 4.5 presents the results of the ablation experiments. In addition to the experiments, we developed interactive web interfaces for the two *GeoJSON Agents* using Python-based frameworks—Streamlit and LangGraph, respectively. As illustrated in Figures 5 and 6, these interfaces allow users to directly upload data and obtain analysis results, thereby further enhancing the usability and practical applicability of the proposed approach.

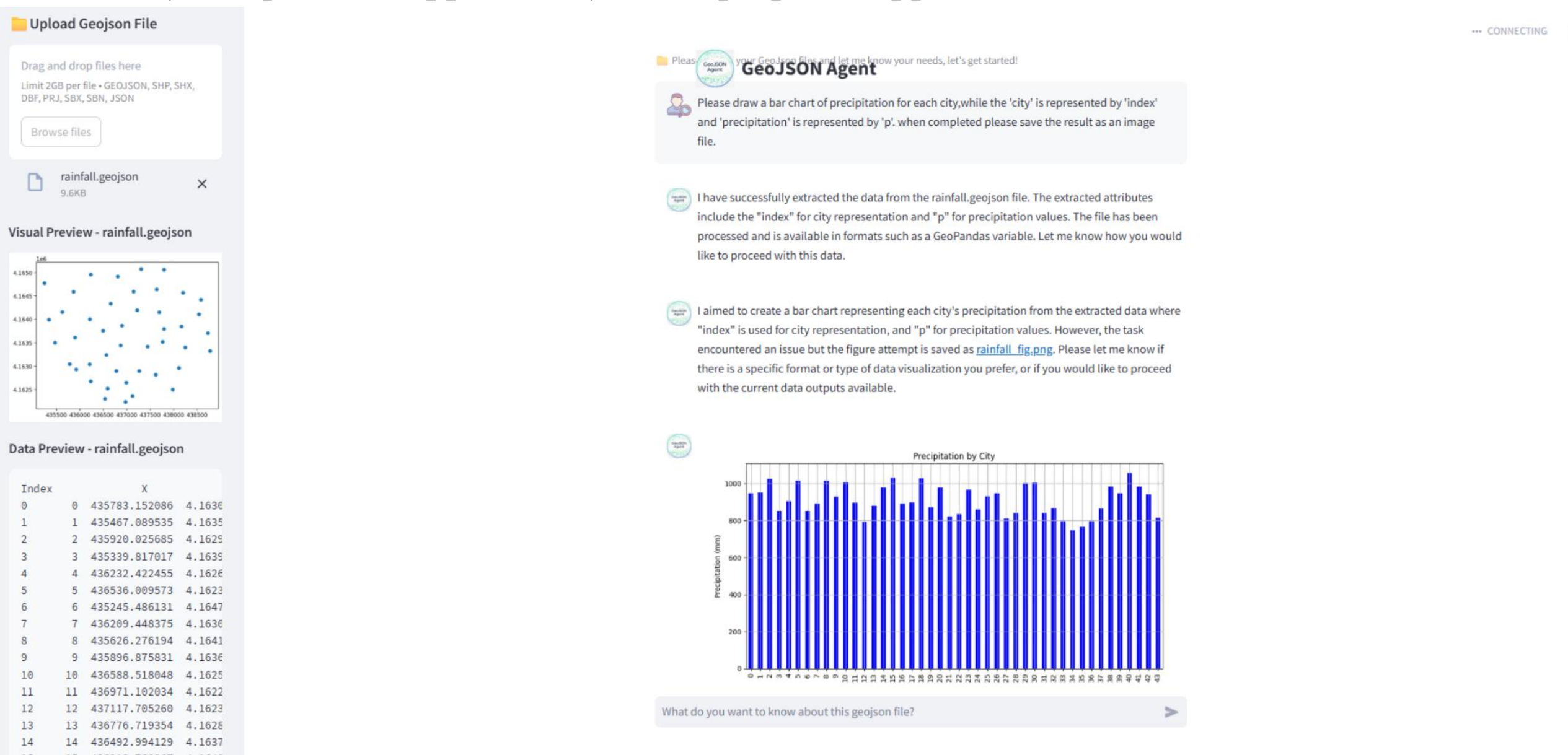

**Figure 5. Interactive web interface of GeoJSON Agent(Function Calling)**, developed using the Streamlit framework. The interface allows users to upload GeoJSON files, submit natural language instructions, and receive processed outputs.

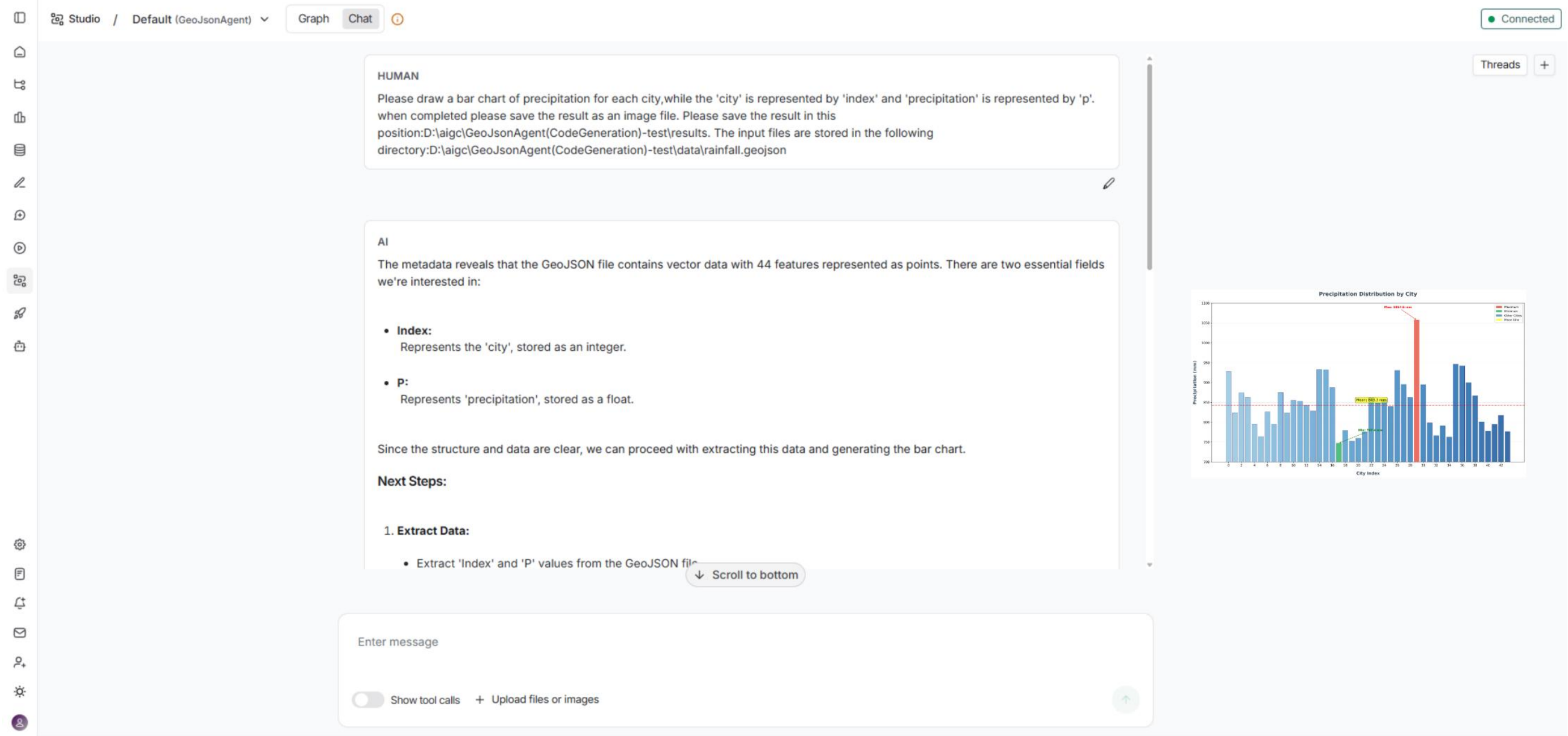

**Figure 6. Interactive web interface of GeoJSON Agent(Code Generation)**, developed using the Langgraph framework. The interface allows users to upload GeoJSON files, submit natural language instructions, and receive processed outputs.

### *4.1. Performance Comparison between GeoJSON Agents and General-Purpose Models*

Using the same set of 70 task cases and corresponding datasets, we conducted a detailed discussion and analysis of the results produced by the two GeoJSON Agents and several general-purpose models (GLM-4-2024-05-20, GLM-4.6-2025-09-30, gpt-4o-mini-2024-07-18, gpt-4o-2024-11-20, and o4-mini-high-2025-04-16). For the general-purpose models—GLM-4-2024-05-20, GLM-4.6-2025-09-30, gpt-4o-mini-2024-07-18, gpt-4o-2024-11-20, and o4-mini-high-2025-04-16—we recorded the generated code and the final output files produced during task execution for each case. These results were used in subsequent analysis to evaluate the models' performance in executing spatial tasks. For the two GeoJSON Agents, we manually reviewed the function call logs or generated code files for each task, as well as the final output after execution, to determine whether the agents produced correct results based on the same task prompts and successfully completed the tasks.

Figure 7 illustrates the number of tasks successfully completed by each model across all datasets. Table 2 further presents the task accuracy and average execution rounds for all models. The Zhipu AI models—GLM-4-2024-05-20 and GLM-4.6-2025-09-30—along with OpenAI's gpt-4o-mini-2024-07-18, demonstrated relatively lower performance, achieving task accuracies of 32.86%, 34.28%, and 34.28%, respectively. In contrast, o4-mini-high-2025-04-16 and gpt-4o-2024-11-20 achieved higher accuracies of 45.71% and 48.57%, showing modest improvement over the previous models. However, both GeoJSON Agents exhibited substantially superior performance on the same tasks. The function-calling-based GeoJSON Agent achieved an accuracy of 85.71%, while the code-generation-based GeoJSON Agent reached an even higher accuracy of 97.14%. Both models significantly outperformed the five general-purpose LLMs across all datasets.

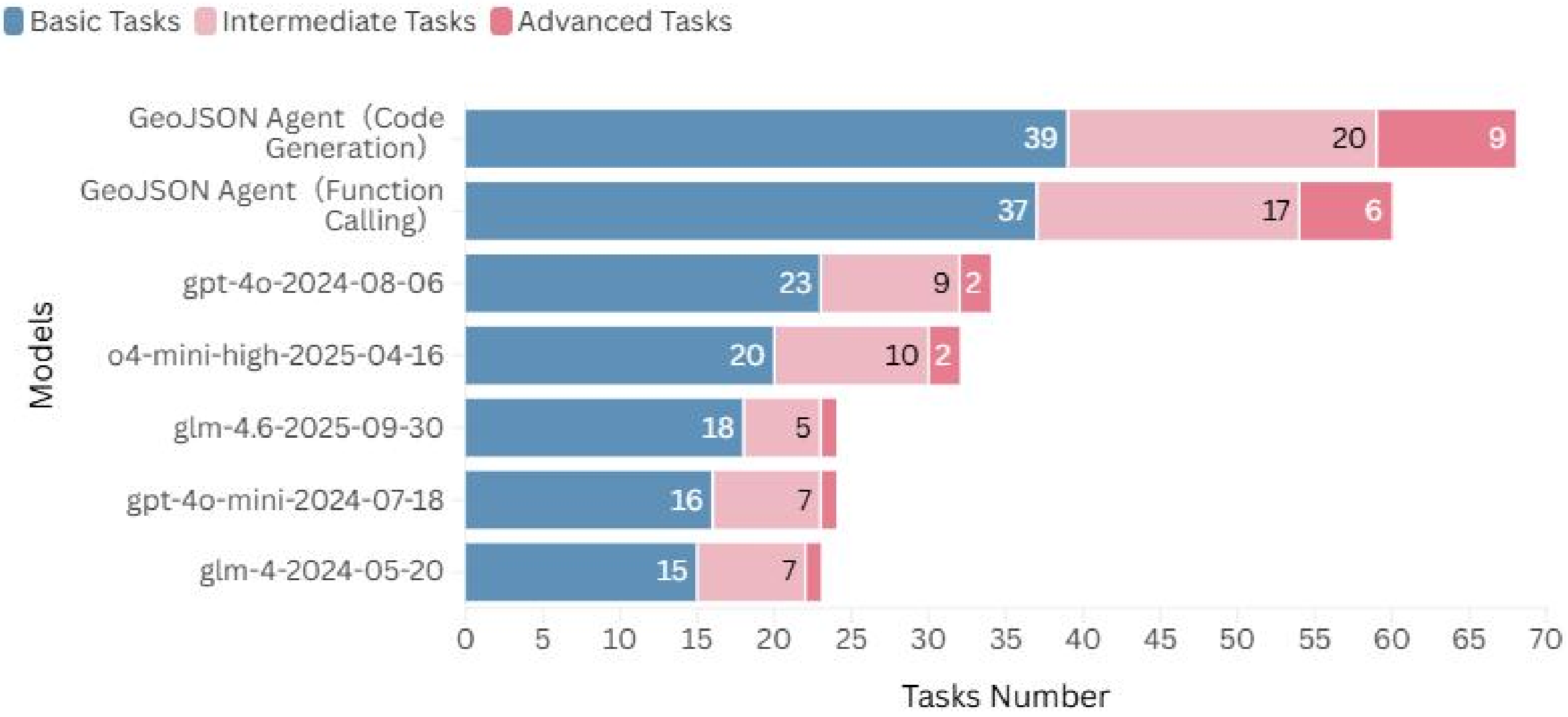

**Figure 7. Task performance comparison between GeoJSON Agents and general models in basic, intermediate, and advanced geospatial tasks.** The results indicate that both GeoJSON Agents outperform general LLMs across all task levels, with the code generation–based agent achieving the highest overall performance, followed by the function-calling–based agent. General models complete far fewer intermediate and advanced tasks, underscoring the advantages of the specialized multi-agent design.

**Table 2. Task performance comparison between multi-agent models and general models.** The results indicate that both GeoJSON Agents outperform general LLMs across all task levels, with the code generation–based agent achieving the highest overall performance, followed by the function-calling–based agent. General models complete far fewer intermediate and advanced tasks, underscoring the advantages of the specialized multi-agent design.

| Models | Accuracy | Average Number Of Execution Rounds |
|---|---|---|
| GLM-4-2024-0520 | 32.86% | 1.15 |
| GLM-4.6-2025-09-30 | 34.28% | 1.70 |
| gpt-4o-mini-2024-07-18 | 34.28% | 1.12 |
| o4-mini-high-2025-04-16 | 45.71% | 1.15 |
| gpt-4o-2024-11-20 | 48.57% | 1.08 |
| Function Calling（gpt-4o-2024-11-20） | **85.71%** | 1.27 |
| Code Generation（gpt-4o-2024-11-20） | **97.14%** | 1.35 |

### *4.2. Performance Analysis of GeoJSON Agents*

#### *4.2.1. Performance of GeoJSON Agents of Different Complexity Tasks*

**GeoJSON Agent Based on Function Calling**

The function-calling-based GeoJSON Agent exhibits a declining performance trend as task complexity increases across spatial analysis tasks(see Table 3). For basic tasks, the function-calling approach demonstrates strong stability, achieving an accuracy of 92.5%, and successfully completing the majority of the tasks. However, as task complexity increases, performance drops significantly: accuracy decreases to 85% for intermediate tasks and further declines to 60% for advanced tasks. This trend suggests that while the function-calling method is effective in handling low- to medium-complexity spatial tasks with clear instructions, it still lacks sufficient capability in natural language understanding and function orchestration when confronted with complex tasks that involve vague instructions, require integrative reasoning, or demand multi-step function compositions.

Moreover, the model's execution efficiency also shows a declining trend as task complexity increases. In terms of average execution rounds, the function-calling method completes most basic tasks in a single round (average: 1.19), indicating strong execution stability. However, for intermediate and advanced tasks, the average number of execution rounds increases to 1.35 and 1.50, respectively. This indicates that as operational complexity grows, the model becomes more prone to errors in function selection and parameter filling—such as failed or incorrect function calls—which lead to execution failures or erroneous results, necessitating retries. The observed increase in execution rounds highlights the function-calling-based GeoJSON Agent's limited adaptability and error resilience in complex scenarios, particularly when managing composite task structures or implicit spatial relationships.

**Table 3. Performance of the function-calling-based GeoJSON Agent across different task complexity levels.** The results show that as the task complexity increases, the accuracy of the model gradually decreases, while the average number of execution rounds gradually increases.

| Task Level | Accuracy | Average Number Of Execution Rounds |
|---|---|---|
| Basic Tasks | 92.5% | 1.19 |
| Intermediate Tasks | 85% | 1.35 |
| Advanced Tasks | 60% | 1.5 |
| Total | 85.71% | 1.27 |

In summary, the function-calling-based GeoJSON Agent demonstrates strong execution efficiency and result stability in simple tasks with clear structure and relatively singular objectives. However, when confronted with complex tasks involving vague instructions or multi-step operations, the model exhibits a dual decline in performance—reduced accuracy and increased execution rounds. These findings suggest that future improvements should focus on enhancing the model's ability to interpret task intent and developing a more fault-tolerant function parameter constraint system, in order to improve the adaptability and overall performance of the function-calling approach in complex spatial task scenarios.

**GeoJSON Agent Based on Code Generation**

The code-generation-based GeoJSON Agent demonstrated strong robustness and adaptability to varying task complexities, maintaining high accuracy across all levels(see Table 4). It achieved an accuracy of 97.5% on basic tasks, a 100% success rate on intermediate tasks, and retained a high accuracy of 90% (9 out of 10) even on the most challenging advanced tasks, showing only a slight decline. These results indicate that the code generation approach offers stronger expressiveness and generalization in mapping natural language instructions to spatial analytical operations. It is better suited to handling complex and non-standard user inputs and analytical requirements.

However, the variation in average execution rounds reveals a degree of instability in the code generation approach during task execution. Specifically, the code-generation-based GeoJSON Agent required an average of 1.28 execution rounds for basic tasks, which increased to 1.35 for intermediate tasks and rose significantly to 1.67 for advanced tasks. Although the overall success rate remained high even for complex tasks, completing these tasks often required multiple rounds of code generation and correction to produce executable results. The underlying cause of this phenomenon may lie in the increasing number of execution steps required for complex tasks, combined with the open-ended and ambiguous nature of user-provided natural language instructions. In such scenarios, the model must autonomously interpret the instruction, decompose it into subtasks, and generate the corresponding code. However, minor misinterpretations can lead to incorrect task decomposition or logical and syntactic errors in the generated code, ultimately resulting in execution failures. Although we incorporated a self-debugging and correction mechanism that allows the model to revise erroneous code within a limited number of attempts (up to five), this mechanism remains insufficient to fully compensate for interpretation errors in highly complex tasks, occasionally resulting in task failure. As a result, the model often requires multiple execution attempts to complete the task successfully, leading to an increased number of execution rounds.

**Table 4. Performance of the code-generation-based GeoJSON Agent across different task complexity levels.** The results demonstrate that the agent achieves near-perfect accuracy on basic and intermediate tasks and maintains strong performance even on advanced tasks, highlighting its robustness and adaptability in handling complex geospatial analyses.

| Task Level | Accuracy | Average Number Of Execution Rounds |
| --- | --- | --- |
| Basic Tasks | 97.5% | 1.28 |
| Intermediate Tasks | 100% | 1.35 |
| Advanced Tasks | 90% | 1.67 |
| Total | 97.14% | 1.35 |

In summary, while the code-generation-based GeoJSON Agent performs exceptionally well in maintaining high accuracy and demonstrates strong adaptability across tasks, its execution stability is comparatively weaker. As task complexity increases, the average number of execution rounds rises noticeably, highlighting the limitations of code generation in maintaining consistent performance during execution. This finding offers valuable insights for future research. Subsequent model designs may benefit from the integration of structured constraints, pre-execution validation, and more fine-grained error diagnosis mechanisms, with the goal of improving the one-pass success rate of generated code, reducing redundant executions, and ultimately enhancing overall system responsiveness and user experience.

*4.2.2. Comparative Analysis of Two Methods Performance and Their Application Scenarios*

Given that the two GeoJSON Agents share an identical architectural design—differing only in their implementation approach—and are evaluated using the same task benchmark set, we conducted a systematic comparison between the function-calling and code-generation implementations. The results are presented in table 5 and figure 8. The evaluation metrics include accuracy and average number of execution rounds, covering basic, intermediate, and advanced task complexities, along with overall aggregated performance.

As shown in the table 5, the code-generation-based GeoJSON Agent demonstrates superior task completion performance across all levels of spatial task complexity. The accuracy gap between the two approaches widens as task difficulty increases. The code-generation-based Agent achieves an overall accuracy of 97.14%, significantly outperforming the function-calling approach, which achieves 85.71%. Notably, it achieved 100% accuracy on intermediate tasks and maintained a high accuracy of 90% on the most challenging advanced tasks—substantially higher than the 60% accuracy of the function-calling method. These results reflect the stronger generalization and flexibility of the code generation approach in both understanding and executing spatial tasks. It demonstrates higher robustness and code synthesis capability, particularly when dealing with structurally complex tasks, ambiguous instructions, and multi-step operations. In contrast, the function-calling approach demonstrates limited adaptability to complex tasks, with accuracy declining sharply as task complexity increases. This exposes a fundamental limitation of function-calling-based agents—their inability to flexibly combine existing functions—resulting in notable deficiencies when dealing with multi-step, complex, and open-ended tasks, as well as in the nuanced interpretation of natural language instructions. However, despite the high task completion rate, the code generation method exhibits lower execution efficiency and stability compared to the function-calling approach. In terms of average execution rounds, the code-generation-based Agent required more rounds across most levels of task complexity. Its overall average was 1.35, whereas the function-calling-based Agent averaged only 1.27. This difference is especially evident in advanced tasks, where the code generation Agent's average execution rounds rose to 1.67, compared to 1.50 for the function-calling model. This indicates that while the code generation method has stronger task completion capabilities, it suffers from weaker structural constraints. Lacking explicit function boundaries, parameter validation, and type-checking mechanisms, the generated code is more susceptible to syntax errors, logical flaws, or environmental dependencies, which often result in execution failures. As a result, the model frequently requires multiple attempts to produce executable code. Although a self-debugging and repair mechanism was incorporated—allowing the model to attempt corrections based on error feedback within a maximum of five rounds—this strategy is often insufficient for complex error cases, where the model must undergo multiple cycles of generation and revision before producing syntactically and logically correct code that meets execution requirements. In contrast, the function-calling-based GeoJSON Agent performs better in terms of average execution rounds, likely due to the inherently structured nature of the method. With predefined function pathways and clear execution logic, the model only needs to identify and invoke the correct functions in sequence to achieve the desired outcome—without having to generate code manually for each subtask—thereby avoiding many of the issues associated with unconstrained code generation.

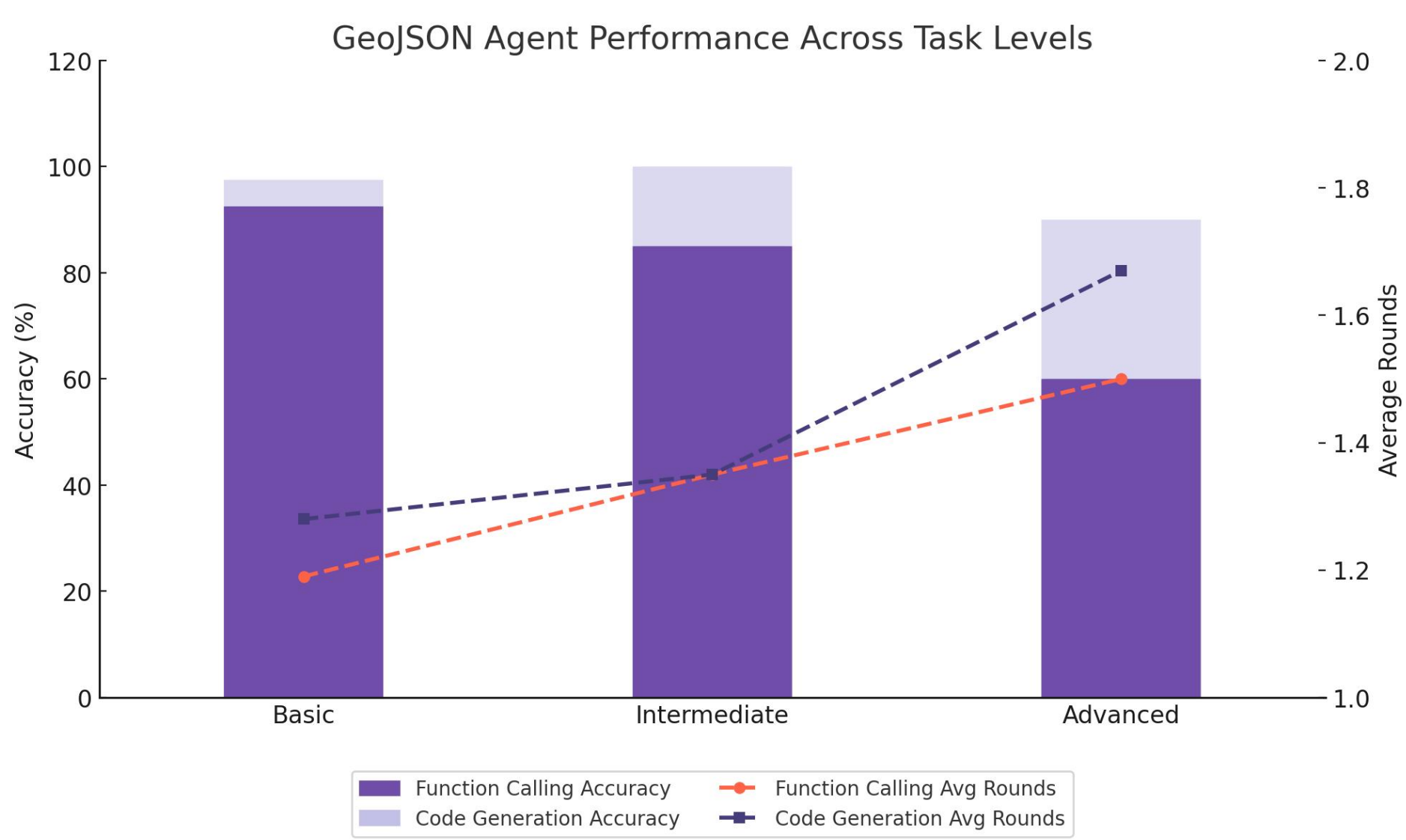

**Figure 8. Performance comparison of the two GeoJSON Agents across task complexity levels.** The Code Generation-based GeoJSON Agent achieves higher accuracy across all complexity levels, while the Function Calling-based GeoJSON Agent exhibits greater execution stability.

Based on the above analysis, it is evident that the function-calling and code-generation approaches each offer distinct advantages in GIS spatial analysis tasks and are suited to different application scenarios. The function-calling method exhibits strong predictability and reliability. Within its functional scope—particularly for low to medium complexity tasks—it can consistently produce high-quality results in a stable and efficient manner. Owing to its structured design and reliance on predefined function interfaces, this approach is better suited for tasks with clear semantics, well-defined objectives, and straightforward operational steps. For instance, in standardized operations such as buffer analysis, area calculation, and spatial filtering, the function-calling approach can efficiently identify the required functions and execute them accurately. It offers high execution efficiency and result reliability, making it particularly suitable for system environments where stability, safety, and performance are critical, and where workflows are fixed and well-structured. In contrast, the code-generation approach demonstrates stronger task generalization capability and greater expressive flexibility. This approach can dynamically generate executable code in response to natural language instructions, making it well-suited for spatial analysis scenarios involving complex task structures, open-ended semantics, or uncertain operational pipelines—especially for problems that are exploratory or creative in nature. For example, when addressing tasks such as “Identify residential areas within 1 km of schools that meet green space coverage requirements”—which are highly integrative and lack explicit operational instructions—the code generation method can autonomously infer intent, decompose the task, and construct a complete analytical workflow, producing a highly customized code solution. Compared with the function-calling approach, it exhibits a distinct advantage in handling open-ended problems.

**Table 5. Comparative performance of the two GeoJSON Agents across task complexity levels.** The results show that the code-generation-based GeoJSON agent performs better on all complex tasks, especially in advanced tasks, but it has a higher average number of execution rounds and poorer stability.

| Task Level | GeoJSON Agent（Function Calling） | | GeoJSON Agent（Code Generation） | |
|---|---|---|---|---|
| | Accuracy | Average Number Of Execution Rounds | Accuracy | Average Number Of Execution Rounds |
| Basic Tasks | 92.5% | 1.19 | 97.5% | 1.28 |
| Intermediate Tasks | 85% | 1.35 | 100% | 1.35 |
| Advanced Tasks | 60% | 1.50 | 90% | 1.67 |
| Total | 85.71% | 1.27 | 97.14% | 1.35 |

### *4.3. Case Study*

To further evaluate the practical capabilities of the two GeoJSON Agents in spatial analysis tasks—and to gain deeper insights into their performance differences and applicability across varying levels of task complexity—this section analyzes three representative case studies selected from basic, intermediate, and advanced task categories. Through these cases, we aim to intuitively demonstrate the effectiveness of both Agents, as well as the real collaborative and communicative mechanisms between the Planner and Worker within the multi-agent architecture, including task decomposition, instruction transmission, state synchronization, and error correction processes. Each case not only presents the complete workflow of natural language instructions, subtask generation, and execution (e.g., function call sequences or code-generation fragments) but also records the multi-turn communication and feedback between the two agents, illustrating the system's dynamic collaboration and conflict-resolution capabilities when addressing complex tasks. Collectively, these cases clearly reveal how multi-agent collaboration performs in handling GIS tasks of varying complexity.

#### *4.3.1. Basic Task: Simple Buffer Analysis*

This case involves generating a 100-meter buffer around a given polygon feature, representing a basic spatial analysis task containing only a single subtask. The input data consist of a polygonal GeoJSON vector dataset provided by the user. The execution process clearly illustrates the division of labor and collaboration between the Planner and Worker (as shown in Figure 9). During task execution, in both approaches, the Planner first decomposes the user's request into the subtask "create a 100-meter buffer" and generates the corresponding execution instruction, which is then passed to the Worker for implementation. Upon receiving the instruction, the Worker focuses solely on execution. In the function-calling-based GeoJSON Agent (Figure 9), the Worker identifies and invokes the appropriate buffer-construction function, CreateMultiRingBufferFromGeoDataFrame, from the predefined function library, successfully producing the target spatial dataset. This process, relying on existing functional components, is computationally efficient and demonstrates the stability and reliability of the function-calling approach for simple tasks. In contrast, within the code-generation-based GeoJSON Agent (Figure 8), the Worker dynamically generates and executes the corresponding Python code in the runtime environment, completing the entire workflow—from data loading and buffer creation to result saving. The final output is successfully stored in the designated directory, highlighting the high flexibility and generative capacity of the code-generation approach. Throughout this process, collaboration between agents is embodied in the feedback loop between the Planner and the Worker. As illustrated in Worker Step 1 of Figure 9, after successful execution, the Worker must report its completion status back to the Planner. Upon receiving this success feedback, the Planner verifies task closure and finalizes the response to the user.

In summary, for simple tasks with clear objectives and single-step processes, the multi-agent collaboration workflow—planning → execution → observation → replanning—demonstrates a high degree of reliability. Both approaches show stable performance: the function-calling method achieves efficient and consistent execution through modular function invocation, while the code-generation method attains flexible goal completion through dynamic code creation. In both cases, success fundamentally depends on the Planner’s accurate task decomposition and the Worker’s precise execution and feedback. This indicates that both approaches exhibit strong and stable performance when handling basic-level tasks.

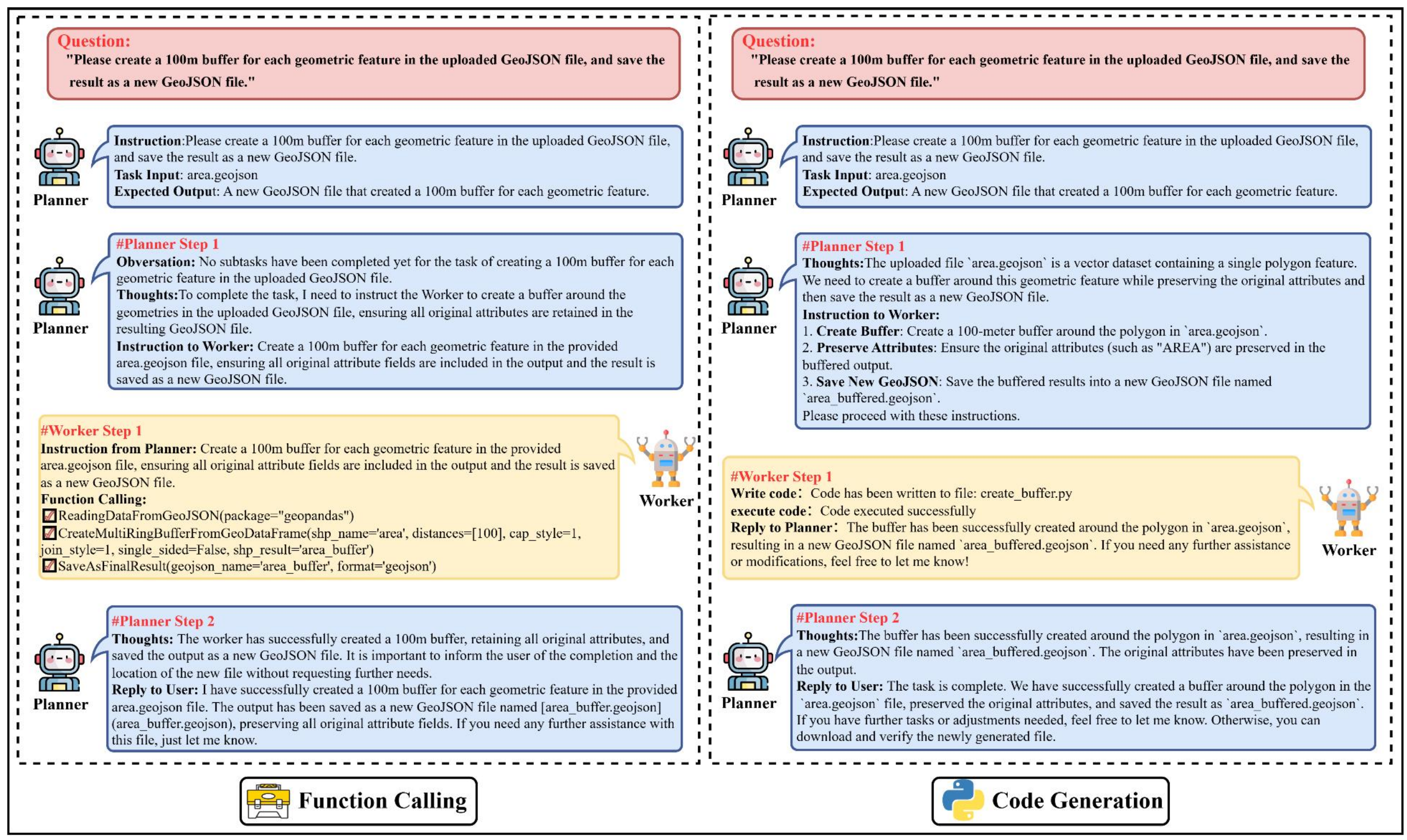

**Figure 9. Workflow comparison of GeoJSON Agents using function calling and code generation in basic simple buffer analysis task**, demonstrates the division of labor and

collaboration between the Planner and Worker. The comparison shows that both agents successfully completed the task.

### *4.3.2. Intermediate Task: Overlap Analysis of Service Areas Between Cafés and Bus Stops*

This case involves creating 500-meter buffer zones around each coffee shop and bus stop, analyzing their overlapping spatial areas, and saving the results as a new Shapefile. The input data consist of user-provided GeoJSON vector datasets containing all coffee shops and bus stops in New York City. During task execution, the Planners of both GeoJSON Agents were responsible for interpreting the user's request, but they adopted distinct task decomposition strategies. In the function-calling-based GeoJSON Agent (Figure 10), the Planner decomposed the user instruction into four detailed subtasks: (1) create 500-meter buffers for all coffee shops in New York City; (2) create 500-meter buffers for all bus stops; (3) analyze the spatial intersection between the two buffer layers; and (4) save the overlapping results as a new Shapefile. The Planner sequentially transmitted each subtask to the Worker, which in turn executed the corresponding functions— "CreateMultiRingBufferFromGeoDataFrame", "OverlayAnalysis", and "SaveAsFinalResult"—while reporting execution feedback after each step. The Planner proceeded with the next subtask only after confirming successful completion of the previous one. This stepwise coordination ensured stable and reliable execution of the multi-step workflow. By contrast, the code-generation-based GeoJSON Agent employed a more compact decomposition strategy (Figure 10). The Planner divided the workflow into three subtasks: (1) create 500-meter buffers for both coffee shops and bus stops; (2) perform spatial overlap analysis; and (3) save the final output. Although the Worker encountered a KeyError during the second subtask, its built-in self-repair and debugging mechanism was activated, allowing it to autonomously diagnose, correct, and re-execute the faulty code. The Worker ultimately reported successful execution to the Planner. The code-generation-based GeoJSON Agent not only successfully fulfilled the user's complex request but also demonstrated strong capabilities in autonomous code debugging and self-correction.

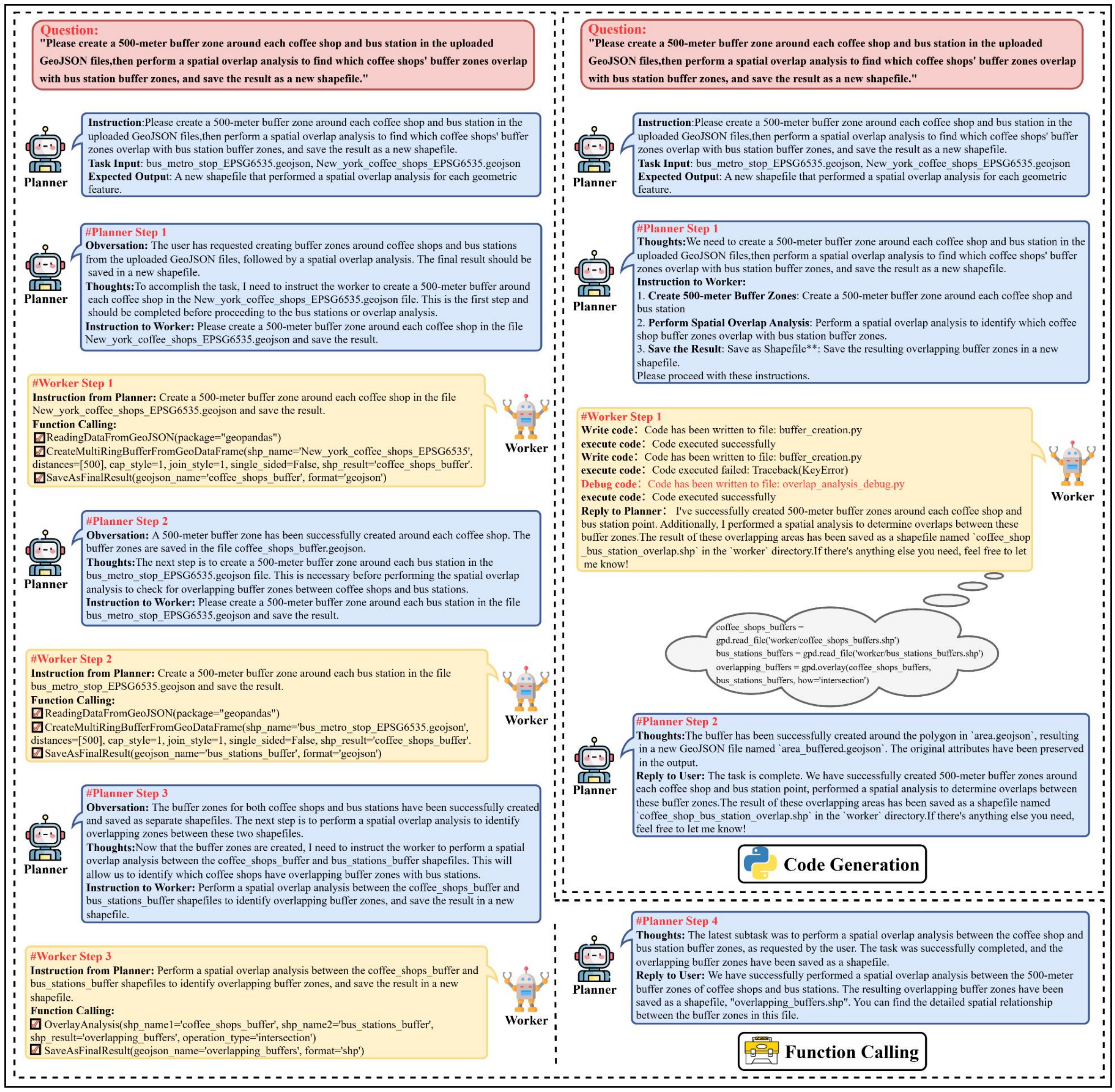

**Figure 10. Workflow comparison of GeoJSON Agents using function calling and code generation in intermediate overlap analysis task.**

In summary, when dealing with multi-step and relatively complex spatial analysis tasks, both the function-calling and code-generation approaches within the Planner–Worker multi-agent architecture were able to perform successfully, demonstrating strong adaptability and execution capability. The function-calling method achieved a stable and reliable workflow through well-defined task division and orderly coordination, leveraging existing functions in a structured manner. In contrast, the code-generation method exhibited greater collaborative efficiency and dynamic adaptability. The Planner was able to operate at a higher level of abstraction, integrating and simplifying task planning, while the Worker served as a robust executor. Its built-in self-repair and debugging mechanisms significantly enhanced the system's robustness and flexibility, underscoring its key advantages in handling complex spatial analysis tasks.

#### *4.3.3. Advanced Task: Generating a Population Density Thematic Map*

This case involves generating a population density map for all counties in Pennsylvania based on existing census data. The input data consist of user-provided GeoJSON vector datasets representing the population census data of each county in the state. This is an advanced, semantically ambiguous, and open-ended task. In this scenario, the two approaches displayed notable differences in their multi-agent collaboration processes. As shown in Figure 11, within the function-calling-based GeoJSON Agent, the Planner successfully decomposed the user's request into several procedural steps, including data loading, field addition, and field calculation. The collaboration between the Planner and Worker proceeded smoothly in the early stages: the Planner sequentially issued subtasks for data loading and field creation, and the Worker accordingly invoked the functions "ReadingDataFromGeoJSON", "SaveAsFinalResult", and "AddFieldToGeoDataFrame", successfully adding a new field named population_density to the dataset and reporting its progress back to the Planner. However, at the critical stage of attribute computation, although the Planner correctly issued the instruction for calculating the field values, the Worker mistakenly invoked the function "UpdateDynamicAttributes" instead of the intended field-calculation function. The Worker reported this execution error to the Planner, which attempted to replan the process but ultimately failed to complete the task. In contrast, when faced with this more complex and open-ended task, the code-generation-based GeoJSON Agent exhibited significantly stronger dynamic collaboration capabilities. As shown in Figure 11, given the ambiguity of the task description, the Planner adopted a more sophisticated decomposition strategy. Instead of immediately breaking down the task, it first issued an exploratory subtask—data analysis. After completing this step, the Worker reported to the Planner the key field information required for population density computation (ALAND and Population). Based on this feedback, the Planner dynamically formulated two concrete subtasks: "calculate density" and "generate map." The Worker then autonomously generated and executed Python code according to these instructions, successfully completing both attribute computation and map generation. This adaptive and iterative workflow enabled the entire complex process to be executed successfully.

In summary, for complex and advanced spatial analysis tasks, the function-calling and code-generation approaches exhibit substantial differences. Although the function-calling method performs reasonably well in task interpretation and basic operations, it fails at critical computational stages due to function invocation errors, revealing its limitations in function matching and error recovery. In contrast, the code-generation approach not only conducts exploratory analyses to actively identify essential fields but also demonstrates greater task flexibility and contextual adaptability. It can dynamically generate executable code and handle complex logic according to task requirements. This case further validates the code-generation method's significant advantages in managing complex and open-ended spatial analysis tasks—particularly in scenarios where workflows cannot be predefined—showcasing its superior flexibility and scalability.

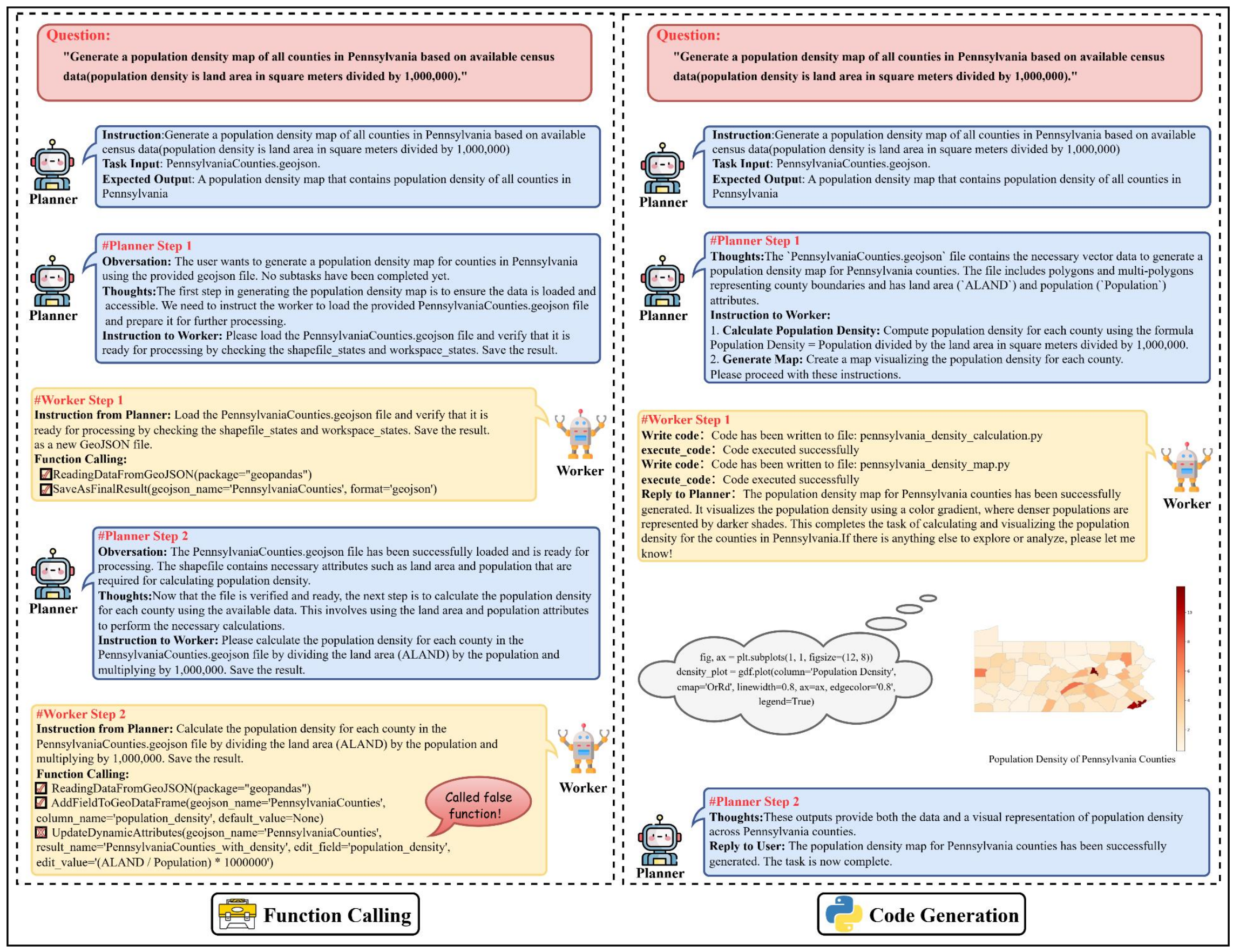

**Figure 11. Workflow comparison of GeoJSON Agents using function calling and code generation in advanced population density thematic map task.** The comparison reveals that while the function-calling method encountered execution failures in complex attribute computations, the code-generation approach dynamically adapted through exploratory analysis and re-planning, successfully completing the full mapping workflow.

### *4.4.Comparative Analysis of GeoJSON Agents with Different Configurations*

This section presents a systematic comparison of the performance of the two GeoJSON Agent architectures under different LLM configurations, aiming to assess how the underlying model capabilities influence spatial task execution. By analyzing experimental results across various Planner–Worker combinations, this study seeks to identify the strengths and weaknesses of each configuration in spatial tasks, thereby providing performance boundary references for model selection and system deployment of the GeoJSON Agents.

In the system architecture of the GeoJSON Agents, each agent is uniformly divided into two key modules: the Planner and the Worker. These two components have clearly defined responsibilities and work collaboratively to fulfill user requests. The Planner is responsible for interpreting user requests, decomposing them into subtasks, and generating clear execution instructions. The Worker, in turn, carries out the actual tasks by either invoking predefined functions from the system's function library or automatically generating executable code based on the instructions received. Given their distinct roles, the underlying model capabilities required by the Planner and Worker also differ. The Planner requires strong reasoning and task planning capabilities to ensure the logical coherence and clarity of the decomposed tasks. The Worker, on the other hand, depends on accurate instruction interpretation, contextual consistency, and stable function execution or code generation to ensure the completeness and correctness of task execution.

As shown in Table 6, within the function-calling-based GeoJSON Agent, overall system performance improves significantly with the enhancement of the underlying LLM capabilities. The model's accuracy increases steadily, while the average number of execution rounds decreases, reflecting the synergistic advantages of stronger models in task comprehension, instruction generation, and function invocation. The first configuration—using gpt-4o-mini-2024-07-18 as both the Planner and the Worker—achieved an accuracy of 77.14% and an average of 2.43 execution rounds, serving as the performance baseline. In the second configuration, upgrading the Planner to gpt-4o-2024-11-20 increased accuracy to 81.43% and reduced the average execution rounds to 2.19, indicating that a more capable Planner can better interpret tasks and generate higher-quality instructions, thereby reducing the execution burden on the Worker. In the third configuration, the Worker was also upgraded to gpt-4o-2024-11-20, resulting in an accuracy of 85.71% and a substantial reduction in average execution rounds to 1.27—the best outcome among all configurations. This demonstrates that a high-performing Worker not only interprets Planner instructions more accurately but also executes functions more effectively, significantly improving task completion efficiency and system stability. The fourth configuration employed OpenAI’s o4-mini-high-2025-04-16 model, released in the first half of 2025, achieving an accuracy of 80% and an average of 1.88 execution rounds. Notably, although o4-mini-high-2025-04-16 demonstrates enhanced reasoning capabilities compared with gpt-4o-mini-2024-07-18, its accuracy improved by only about 3%, indicating relatively limited performance gains. This outcome can be attributed to the model’s reasoning-oriented design: when handling function-calling tasks, it tends to produce longer reasoning chains and more detailed contextual information. In the multi-turn interaction setting of GeoJSON Agents, this behavior leads to rapid accumulation of contextual length, which in turn compromises the model’s stability in subsequent execution rounds. This finding highlights potential adaptability challenges faced by reasoning-oriented models in specific task scenarios.

**Table 6. Performance of function-calling-based GeoJSON Agents under different configurations.** The results show that the best performance was achieved when both Planner and Worker used the GPT-4o-2024-08-06 model, reaching 85.71% accuracy with the fewer execution rounds.

| ID | Planner | Worker | Accuracy | Average Number Of Execution Rounds |
|---|---|---|---|---|
| 1 | gpt-4o-mini-2024-07-18 | gpt-4o-mini-2024-07-18 | 77.14% | 2.43 |
| 2 | gpt-4o-2024-11-20 | gpt-4o-mini-2024-07-18 | 81.43% | 2.19 |
| 3 | gpt-4o-2024-11-20 | gpt-4o-2024-11-20 | **85.71%** | **1.27** |
| 4 | o4-mini-high-2025- | o4-mini-high-2025- | 80% | 1.88 |

| ID | Planner | Worker | Accuracy | Average Number Of Execution Rounds |
|---|---|---|---|---|
| | 04-16 | 04-16 | | |
| 5 | GLM-4-2024-0520 | GLM-4-2024-0520 | 65.71% | 1.52 |
| 6 | GLM-4.6-2025-09-30 | GLM-4.6-2025-09-30 | 72.86% | 1.21 |
| 7 | qwen-plus-2025-01-25 | qwen-plus-2025-01-25 | 71.43% | 1.78 |

**Table 7. Performance of code-generation-based GeoJSON Agents under different configurations.** The results show that the best performance was achieved when both Planner and Worker used the GPT-4o-2024-08-06 model, reaching 97.14% accuracy with the fewer execution rounds.

| ID | Planner | Worker | Accuracy | Average Number Of Execution Rounds |
|---|---|---|---|---|
| 1 | gpt-4o-mini-2024-07-18 | gpt-4o-mini-2024-07-18 | 94.29% | 1.65 |
| 2 | gpt-4o-2024-11-20 | gpt-4o-mini-2024-07-18 | 97.14% | 1.37 |
| 3 | gpt-4o-2024-11-20 | gpt-4o-2024-11-20 | **97.14%** | **1.35** |
| 4 | o4-mini-high-2025-04-16 | o4-mini-high-2025-04-16 | 92.86% | 1.32 |
| 5 | GLM-4-2024-0520 | GLM-4-2024-0520 | 67.14% | 1.96 |
| 6 | GLM-4.6-2025-09-30 | GLM-4.6-2025-09-30 | 77.14% | 1.52 |
| 7 | qwen-plus-2025-01-25 | qwen-plus-2025-01-25 | 82.86% | 1.91 |

In the code-generation-based GeoJSON Agent, changes in model configuration likewise had a significant impact on system performance. As shown in Table 7, with increasing model capability, task accuracy continued to improve, while the average number of execution rounds consistently decreased. The first configuration, using gpt-4o-mini-2024-07-18 for both the Planner and the Worker, achieved 94.29% accuracy with an average of 1.65 execution rounds, demonstrating solid stability. In the second configuration, replacing the Planner with gpt-4o-2024-11-20 further increased accuracy to 97.14% and reduced the average execution rounds to 1.37, indicating that a more capable Planner contributes to clearer task decomposition and higher-quality code instruction generation. In the third configuration, after upgrading the Worker to the same higher-tier model, the accuracy remained unchanged, but the average execution rounds further dropped to 1.35, highlighting the notable advantages of a high-performance Worker in code generation, error handling, and debugging. This configuration achieved the highest accuracy across all tests, setting a new performance benchmark. The fourth configuration, using o4-mini-high-2025-04-16, achieved an accuracy of 92.86% and an average of 1.32 execution rounds. Surprisingly, although o4-mini-high-2025-04-16 possesses stronger reasoning capabilities in theory than gpt-4o-mini-2024-07-18, its accuracy (92.86%) was slightly lower than that of the first configuration (94.29%) and notably lower than that of gpt-4o-2024-11-20 (97.14%). This pattern mirrors the phenomenon observed in the function-calling experiments and can largely be attributed to the reasoning-oriented characteristics of o4-mini-high-2025-04-16. In code-generation scenarios, this model tends to produce more elaborate reasoning processes and excessively detailed code annotations, which rapidly inflate the context length. During multi-turn iterations and self-debugging cycles, this accumulated context effect distracts the model from the core task logic, reducing both code-generation precision and execution efficiency. This finding further confirms the potential trade-offs that reasoning-oriented models may face in long-context, multi-round interaction tasks.

In addition to OpenAI's models, to comprehensively evaluate the adaptability of GeoJSON Agents across different vendors and model generations, we also tested other mainstream large language models. For both methodological configurations, we incorporated Zhipu AI's GLM-4-2024-05-20 and its updated version GLM-4.6-2025-09-30 as the fifth and sixth comparative groups, respectively, and introduced Alibaba's qwen-plus-2025-01-25 as the seventh comparison group. In the function-calling configuration, the fifth experiment using GLM-4-2024-05-20 achieved an accuracy of 65.71%, lower than the OpenAI models, but with an average execution round of 1.52—better than some OpenAI configurations—indicating relatively high task response efficiency. However, this model still showed limitations in function matching and API-calling capabilities, making it less reliable for complex spatial tasks. The sixth configuration, using the updated GLM-4.6-2025-09-30, achieved 72.86% accuracy and 1.21 average execution rounds. Although still below the OpenAI models, its performance improved notably compared to GLM-4-2024-05-20, reflecting enhanced reliability and progress in function matching and API utilization. This demonstrates the positive impact of continual model optimization on improving automation in GIS-related tasks. The seventh configuration, using qwen-plus-2025-01-25, achieved 71.43% accuracy and 1.78 average execution rounds, suggesting that the Qwen series also possesses solid baseline capabilities for function-calling tasks, with performance comparable to the updated GLM-4.6, though still trailing the OpenAI gpt models. In the code-generation configuration, the fifth experiment with GLM-4-2024-05-20 achieved an accuracy of 67.14% and an average of 1.96 execution rounds, reflecting modest performance with basic task execution capability. The sixth configuration, employing the updated GLM-4.6-2025-09-30, improved to 77.14% accuracy and 1.52 average execution rounds. Compared with GLM-4-2024-05-20, this represents a clear enhancement, indicating that the updated GLM model made tangible progress in semantic understanding and self-debugging abilities, showing a performance improvement trend consistent with the function-calling results. Nevertheless, its accuracy remained below that of the gpt series, suggesting that while the GLM family can effectively generate and execute basic logical code, it still lags behind in semantic comprehension, multi-step reasoning, and error recovery. The seventh configuration using qwen-plus-2025-01-25 achieved an accuracy of 82.86% and an average of 1.91 execution rounds, significantly outperforming the contemporary GLM-4.6-2025-09-30 and exhibiting stronger competitiveness in code-generation tasks. However, its performance still fell short of the gpt series, indicating room for further optimization.

Overall, the experimental results indicate that more powerful foundational models can significantly enhance the GeoJSON Agent's ability to understand spatial tasks, improve execution efficiency, and increase output accuracy—effects that are particularly pronounced in high-complexity task scenarios. At the same time, the GeoJSON Agent architecture also demonstrates a degree of generalizability and transferability when applied to non-OpenAI models, suggesting its feasibility for deployment in multi-model environments. Furthermore, the continuous evolution and updates of the same model family have had a positive impact on performance optimization. For instance, the GLM series demonstrated significant improvements from GLM-4-2024-05-20 to GLM-4.6-2025-09-30, with accuracy gains of 7% and 10% in the function-calling and code-generation configurations, respectively, thereby enhancing the overall automation level of GIS tasks. However, not all more capable models achieved the expected performance improvements within the GeoJSON Agent framework. The reasoning-oriented o4-mini-high-2025-04-16, released by OpenAI in 2025, performed strongly on standard benchmarks but yielded only marginal—or even reduced—performance gains in our multi-turn, long-context GeoJSON Agent setting. This finding suggests that when selecting LLM configurations, one must consider not only a model's overall capability but also its alignment with specific application scenarios. For tasks requiring concise outputs and efficient execution, traditional generative models may, in fact, be more suitable than reasoning-oriented ones. Future research may further explore instruction tuning, API adaptation, and task decomposition strategies tailored to different LLM architectures, with the goal of enhancing the system's adaptability and performance across diverse model ecosystems.

### *4.5.Ablation Study*

In this section, to quantitatively evaluate the necessity and contribution of the Planner agent module within the GeoJSON Agents architecture—and to assess its effectiveness in guiding the Worker agent module—we conducted an ablation study on the Planner. The core of this experiment lies in comparing the performance of the complete "Planner + Worker" multi-agent architecture with that of a single-agent configuration in which the Planner is removed and only the Worker remains. In the single-agent configuration, the Worker directly receives the user's natural language instructions and independently performs all processes, including task understanding, step decomposition, and final execution. This ablation experiment was conducted on both GeoJSON Agent implementations—the function-calling and code-generation variants—using two LLM configurations with different capacities (gpt-4o-2024-08-06 and gpt-4o-mini-2024-07-18) as the Worker's underlying model. The analysis focused particularly on changes in accuracy and average execution rounds to comprehensively assess the Planner module's value. Tables 8 and 9 summarize the experimental results under each configuration.

**Table 8. Ablation experiment results of the Planner module in the function-calling-based GeoJSON Agents.** The results demonstrate that removing the Planner module leads to a substantial performance drop—accuracy decreased from 85.71% to as low as 12.86%—highlighting the critical role of the Planner in task decomposition, coordination, and overall system reliability.

| ID | Planner | Worker | Accuracy | Average Number Of Execution Rounds |
|---|---|---|---|---|
| 1 | gpt-4o-2024-11-20 | gpt-4o-2024-11-20 | 85.71% | 1.27 |

| ID | Planner | Worker | Accuracy | Average Number Of Execution Rounds |
|---|---|---|---|---|
| 2 | × | gpt-4o-2024-11-20 | **45.71%** | 1.47 |
| 3 | gpt-4o-2024-11-20 | gpt-4o-mini-2024-07-18 | 81.43% | 2.19 |
| 4 | gpt-4o-mini-2024-07-18 | gpt-4o-mini-2024-07-18 | 77.14% | 2.43 |
| 5 | × | gpt-4o-mini-2024-07-18 | **12.86%** | 1.33 |

**Table 9. Ablation Experiment Results of the Planner Module in the code-generation-based GeoJSON Agents.** The results show that removing the Planner module causes a notable performance degradation, with accuracy dropping from 97.14% to 40–58.57%, demonstrating that the Planner is essential for maintaining high accuracy and effective coordination in complex geospatial reasoning tasks.

| ID | Planner | Worker | Accuracy | Average Number Of Execution Rounds |
|---|---|---|---|---|
| 1 | gpt-4o-2024-11-20 | gpt-4o-2024-11-20 | 97.14% | 1.35 |
| 2 | × | gpt-4o-2024-11-20 | **58.57%** | 1.56 |
| 3 | gpt-4o-2024-11-20 | gpt-4o-mini-2024-07-18 | 97.14% | 1.37 |
| 4 | gpt-4o-mini-2024-07-18 | gpt-4o-mini-2024-07-18 | 94.29% | 1.65 |
| 5 | × | gpt-4o-mini-2024-07-18 | **40%** | 1.43 |

In the function-calling–based GeoJSON Agent, when both the Planner and Worker were configured with gpt-4o-2024-08-06 (Configuration 1), the overall system performance was strong, achieving an accuracy of 85.71% and an average of 1.27 execution rounds. However, when the Planner was removed from Configuration 1 (Configuration 2), the accuracy dropped sharply to 45.71%, while the average execution rounds increased to 1.47. Even with a high-performing Worker model (gpt-4o-2024-08-06), the absence of a Planner made it difficult for the system to independently perform complex task interpretation and construct coherent chains of function calls. When the Worker's LLM configuration was replaced with a less capable model (gpt-4o-mini-2024-07-18, Configuration 5), the model's accuracy declined even more drastically—from 45.71% to 12.86%. Although the average execution rounds slightly decreased to 1.33, the resulting task outputs were severely distorted, indicating that weaker single-agent configurations are almost incapable of autonomously completing geospatial tasks. When Planners were reintroduced—gpt-4o-mini-2024-07-18 (Configuration 4) and gpt-4o-2024-08-06 (Configuration 3)—the accuracy rose significantly to 77.14% and 81.43%, respectively. This result strongly demonstrates the crucial role of the Planner module: even when the Worker is relatively limited, a capable Planner can markedly enhance overall system performance by providing clear task decomposition, precise function-matching guidance, and well-structured execution planning. In contrast, removing the Planner leads to disorganized instruction logic and incomplete function-call sequences, resulting in substantial drops in both accuracy and execution stability. The Planner thus plays an essential role in decomposing tasks, coordinating execution, and guiding the Worker toward successful task completion.

A similar trend was observed in the code-generation–based GeoJSON Agent, although the performance degradation was even more pronounced—further underscoring the critical role of the Planner module in this context. When both the Planner and Worker were configured with gpt-4o-2024-08-06 (Configuration 1), the system achieved optimal performance, reaching an accuracy of 97.14% and an average of 1.35 execution rounds. However, after the Planner was removed (Configuration 2), despite the Worker still using the high-performance gpt-4o-2024-08-06 model, the accuracy dropped sharply to 58.57%, and the average execution rounds increased to 1.56. An even steeper decline occurred in Configuration 5, where the Worker used the less capable gpt-4o-mini-2024-07-18. In this single-agent setup, the model's accuracy fell to only 40%, with an average of 1.43 execution rounds—an 18.57 percentage-point drop compared with Configuration 2. This outcome highlights the inherent complexity of code-generation tasks: when the model's capacity is limited and lacks Planner guidance, the Worker struggles to accurately interpret user intent and frequently makes errors in code logic construction, library function invocation, and debugging, resulting in a significant decline in task quality. When the Planner was reintroduced into Configuration 5, the system's performance improved dramatically. Configurations 4 and 3, which used gpt-4o-mini-2024-07-18 and gpt-4o-2024-08-06 as the Planner, achieved accuracies of 94.29% and 97.14%, respectively, while the average execution rounds decreased to 1.37. These results clearly demonstrate the Planner's central role in code-generation tasks: by decomposing complex geospatial problems into clear, executable subtask sequences and providing explicit guidance for code generation, the Planner effectively reduces the Worker's cognitive load, enabling it to focus on producing and refining high-quality code—thereby substantially improving overall system performance.

While the aggregate statistics in Tables 8 and 9 highlight the overall importance of the Planner, a more fine-grained, task-level analysis is required to fully understand the architectural trade-offs. Configurations without a Planner (Configurations 2 and 5 in Tables 8 and 9) essentially function as single-agent systems, in which the Worker directly interprets and executes high-level user instructions. To enable a direct comparison between the single-agent and complete multi-agent architectures, Tables 10 and 11 present the models' performance across tasks of varying complexity. The results clearly show that the Planner's contribution becomes increasingly significant as task complexity rises, a trend consistently observed across both implementation methods. In the function-calling–based GeoJSON Agent (Table 10), the single-agent configuration (Worker: gpt-4o-mini-2024-07-18) achieved only 17.5% accuracy on basic tasks, whereas adding a Planner—gpt-4o-mini-2024-07-18 or gpt-4o-2024-08-06—boosted accuracy to 67.5% and 72.5%, respectively. For intermediate tasks, accuracy in the single-agent setup dropped to 10%, while the multi-agent architecture achieved 75% and 80%. In advanced tasks, the single-agent configuration completely failed (0% accuracy), whereas the multi-agent setup maintained a 50% success rate. The difference was even more striking in the code-generation–based GeoJSON Agent (Table 11): the single-agent system reached 52.5% accuracy on basic tasks, which increased to 95% and 97.5% when the Planner was added. For intermediate tasks, the single-agent configuration achieved 35% accuracy, while the multi-agent system reached 100%. In advanced tasks, the single-agent model managed only 10% accuracy, compared to 80% and 90% with the multi-agent architecture. These results indicate that as task complexity increases, the cognitive load on a single-agent system grows exponentially, leading to severe performance degradation. In contrast, the multi-agent architecture—through the Planner's structured task decomposition and execution guidance—effectively transforms complex problems into a series of manageable, executable subtasks, enabling the system to maintain stable performance even in high-complexity scenarios.

**Table 10. Performance of function-calling-based single-agent and multi-agent models (with different Planner configurations) across task complexity levels.** The results indicate that multi-agent models significantly outperform single-agent models across all task levels, especially in intermediate and advanced tasks, demonstrating the strong advantages of planner in complex spatial reasoning.

| **Task Level** | **Single-Agent (Worker: gpt-4o-mini-2024-07-18)** | **Multi-Agent (Planner: gpt-4o-mini-2024-07-18)** | **Multi-Agent (Planner: gpt-4o-2024-08-06)** |
| --- | --- | --- | --- |
| Basic Tasks | **17.5%** | 85% | 90% |
| Intermediate Tasks | **10%** | 75% | 80% |
| Advanced Tasks | **0%** | 50% | 50% |

**Table 11. Performance of code-generation–based single-agent and multi-agent models (with different Planner configurations) across task complexity levels.** The results clearly show that multi-agent models substantially outperform the single-agent model across all levels of task complexity, highlighting the critical role of the Planner in enhancing coordination, reasoning accuracy, and robustness in complex geospatial analysis.

| **Task Level** | **Single-Agent (Worker: gpt-4o-mini-2024-07-18)** | **Multi-Agent (Planner: gpt-4o-mini-2024-07-18)** | **Multi-Agent (Planner: gpt-4o-2024-08-06)** |
| --- | --- | --- | --- |
| Basic Tasks | **52.5%** | 95% | 97.5% |
| Intermediate Tasks | **35%** | 100% | 100% |
| Advanced Tasks | **10%** | 80% | 90% |

The analysis of the ablation experiments clearly validates the Planner module's pivotal role within the GeoJSON Agents multi-agent architecture. In both the function-calling and code-generation implementations, incorporating the Planner led to substantial performance improvements. Through the coordinated collaboration between the Planner and Worker, the multi-agent mechanism not only significantly enhanced model accuracy and execution efficiency but also strengthened the system's robustness and scalability in handling complex spatial tasks—particularly in medium- and high-complexity scenarios. These findings indicate that the superior performance of GeoJSON

Agents does not merely stem from advances in model capability but rather from the structured intelligence embedded in its multi-agent architecture, which enables effective task planning and coordinated execution. This underscores that well-designed architectural intelligence plays a decisive role in building efficient and reliable geospatial AI systems—its importance even surpassing that of simply employing more powerful models.

## 5. Discussion

### *5.1. Research Contributions and Findings*

This study addresses two key challenges: the high operational barriers of traditional GIS tools and the limited performance of general-purpose large language models (LLMs) in handling complex geospatial tasks. It also responds to the lack of research on GeoJSON data in the GIS domain. To this end, we propose and validate an automated spatial analysis framework based on a multi-agent architecture—GeoJSON Agents. The framework adopts a Planner–Worker architecture that enables clear functional division and cooperative interaction: the Planner is responsible for high-level task understanding, decomposition, and dynamic planning, while the Worker focuses on low-level tool invocation or code generation. Each agent maintains an independent context, prompting strategy, and execution environment, collaborating through explicit message exchange. The framework incorporates two core methodologies—function calling and code generation—designed specifically for processing modern GeoJSON data formats, and was systematically evaluated on a benchmark of 70 tasks with varying levels of complexity. Experimental results show that both GeoJSON Agents significantly outperform general-purpose LLMs. The code-generation-based Agent achieved a task accuracy of 97.14%, and the function-calling-based Agent reached 85.71%, both substantially surpassing the best-performing general LLM, which achieved only 48.57%. Furthermore, results from the ablation experiments revealed that both GeoJSON Agents substantially outperformed their corresponding single-agent models, whose best accuracies were only 45.71% and 58.57%, respectively. These findings validate the effectiveness of the multi-agent architecture in overcoming the limitations of both traditional GIS workflows and general LLMs, and highlight its potential to advance the automation and intelligence of geospatial task execution.

Furthermore, this study provides a comparative analysis of the two methods used within GeoJSON Agents for geospatial analysis tasks, highlighting their respective strengths and limitations. The code-generation-based GeoJSON Agent consistently demonstrated superior performance across varying task complexities—achieving 100% accuracy on intermediate tasks and maintaining 90% accuracy even on complex, open-ended advanced tasks. This success can be attributed to its ability to dynamically generate customized Python scripts using libraries such as GeoPandas, providing the flexibility needed to handle open-ended and complex geospatial tasks. However, the code generation method exhibited a higher average number of execution rounds (1.35) compared to the function-calling method (1.27). This reflects its relative instability, as complex tasks often lead to increased retries due to potential errors in generated code. In contrast, the function-calling-based GeoJSON Agent demonstrated more stable performance in structured tasks, with a lower overall average execution round of 1.27 and higher accuracy in standardized tasks such as buffer analysis and spatial querying. This aligns with findings from Lin et al.'s work on ShapefileGPT, which suggests that function-calling methods exhibit higher reliability in executing predefined and well-constrained tasks. However, our study also addresses a gap in the existing literature by highlighting a key limitation: the function-calling approach exhibits a strong dependence on predefined function libraries, which constrains its performance when handling complex, high-level tasks. Specifically, it struggles to accurately map open-ended or ambiguous user requests to corresponding predefined function-call sequences, resulting in a markedly reduced accuracy of only 60%. This constraint has not been thoroughly examined in previous research.

To comprehensively evaluate the generality and scalability of the GeoJSON Agents framework, this study conducted a systematic comparative analysis across LLMs from different vendors and generations. The evaluation covered OpenAI's GPT series (including gpt-4o-mini-2024-07-18, gpt-4o-2024-08-06, and the newly released o4-mini-high-2025-04-16), Zhipu AI's GLM series (GLM-4-2024-05-20 and GLM-4.6-2025-09-30), and Alibaba's Qwen-plus-2025-01-25. Experimental results demonstrate that improvements in LLM capability can significantly enhance the performance of GeoJSON Agents, though this enhancement is not strictly linear—it depends heavily on the alignment between model characteristics and the target application scenario. For instance, the update from GLM-4-2024-05-20 to GLM-4.6-2025-09-30 yielded substantial accuracy gains in both implementation modes. However, enhanced model capacity does not always translate to expected performance improvements. The reasoning-oriented o4-mini-high-2025-04-16, despite its strong results on standard benchmarks, showed limited or even negative gains in GeoJSON Agents' multi-turn, long-context settings. This is because the model tends to generate lengthy reasoning chains and verbose contextual content, causing rapid context expansion during iterative interactions—thereby distracting the model from the core task logic and reducing both efficiency and accuracy. This finding highlights that model selection should consider not only overall capability but also the compatibility between model characteristics (e.g., reasoning- vs. generation-oriented behavior, long-context handling ability) and the intended application context.

Overall, the proposed GeoJSON Agents framework demonstrates notable advantages in both architectural design and task execution capability, significantly advancing the automation and intelligence of geospatial analysis workflows. By systematically developing a multi-agent model tailored to the GeoJSON format and integrating two mainstream LLM-augmented strategies—function calling and code generation—this study not only validates the applicability and performance differences of each method across multi-level tasks, but also fills a key gap in the GeoAI field regarding the integration of modern data formats with agent-based methodologies. It lays the groundwork for driving GeoAI toward greater efficiency, generality, and intelligence.

### *5.2. Analysis of Token Consumption*

In addition to task accuracy and execution stability, token consumption is also a key indicator for evaluating a model's practical utility, as it directly determines the system's operational cost. This study employed top-tier commercial models such as gpt-4o, whose API invocation cost is relatively high. Under the baseline configuration (Configuration 3: both the Planner and Worker were gpt-4o-2024-08-06), completing all 70 tasks incurred approximately USD 28.22 for the function-calling method and USD 12.26 for the code-generation method, with the latter showing a clear cost advantage. Accordingly, we recorded the average token consumption per task for the two GeoJSON Agents across seven different LLM configurations (see Table 12). A detailed analysis of the token-consumption data reveals that the operational cost differences between the two implementation approaches stem from their fundamentally distinct architectures and prompt-design strategies.

First, in comparing token consumption across implementation methods, we found that the GeoJSON Agent based on the code-generation approach demonstrated a substantial advantage. Under the baseline configuration (Configuration 3: both the Planner and Worker were gpt-4o-2024-08-06), the GeoJSON Agent using the function-calling method consumed approximately 45,000–74,000 tokens per task, whereas the code-generation method required only about 8,100–19,000 tokens. This disparity arises from fundamental differences in prompt-construction strategies. To ensure the accuracy of task planning by the Planner and the correctness of tool selection by the Worker, the function-calling method embeds the complete API function library documentation—containing function names, parameter definitions, usage instructions, and examples (stored in YAML format)—into each round of task planning. Consequently, its contextual input is lengthy and repetitive, leading to significantly higher token consumption. In contrast, the prompts used in the code-generation approach are more concise, including only the task description and essential context. Although the Worker's self-repair and debugging mechanism may trigger multiple iterations (up to five) when handling complex tasks—thereby increasing response time—its total token usage remains considerably lower than that of the function-calling method. This finding underscores the decisive impact of prompt-design strategies on model operating costs: while extensive contextual dependence can improve task accuracy, it does so at the expense of substantially higher token consumption.

**Table 12. Performance of Single-task Token Consumption Range of GeoJSON Agents.** This table compares the single-task token consumption ranges of GeoJSON Agents implemented with function calling and code generation under seven different Planner–Worker configurations. The results reveal significant differences in token efficiency across models and methods.

| ID | Configuration | | Single-task Token Consumption Range | |
|---|---|---|---|---|
| | Planner | Worker | GeoJSON Agent (Function Calling) | GeoJSON Agent (Code Generation) |
| 1 | gpt-4o-mini-2024-07-18 | gpt-4o-mini-2024-07-18 | 45000-81000 | 11000-27000 |
| 2 | gpt-4o-2024-08-06 | gpt-4o-mini-2024-07-18 | 46000-80000 | 11000-26000 |
| 3 | gpt-4o-2024-08-06 | gpt-4o-2024-08-06 | 45000-74000 | 8200-19000 |
| 4 | o4-mini-high-2025-04-16 | o4-mini-high-2025-04-16 | 70000-80000 | 15000-25000 |
| 5 | GLM-4-2024-05-20 | GLM -4-2024-05-20 | 12000-48000 | 9800-13000 |
| 6 | GLM-4.6-2025-09-30 | GLM-4.6-2025-09-30 | 11000-49000 | 9700-12000 |
| 7 | qwen-plus-2025-01-25 | qwen-plus-2025-01-25 | 34000-62000 | 30000-48000 |

Moreover, different LLM configurations have a substantial impact on token consumption. For the GeoJSON Agent based on the function-calling method, token usage showed no significant fluctuation between Configuration 1 (both Planner and Worker using gpt-4o-mini-2024-07-18) and Configuration 3 (both using gpt-4o-2024-08-06). In contrast, for the code-generation method, token consumption decreased slightly from 11,000–27,000 to 8,200–19,000, indicating a trend of improved efficiency. This pattern reflects how LLM configuration influences task-processing efficiency: more capable LLMs tend to converge more quickly toward correct answers during task execution, thereby reducing redundant reasoning. As a result, token consumption decreases in the dynamic code-generation approach, whereas in the more structured function-calling method, model improvements are manifested primarily in task accuracy rather than in token efficiency. Configurations 4 (Planner and Worker both using o4-mini-high-2025-04-16) and 7 (Planner and Worker both using qwen-plus-2025-01-25) exhibited relatively higher token consumption in both implementations, likely due to their respective model characteristics. The o4-mini-high-2025-04-16 model, being inference-oriented, tends to generate longer reasoning chains and more detailed contextual information, causing rapid context accumulation and, consequently, higher token usage. Conversely, qwen-plus-2025-01-25 may require more interaction rounds and longer contexts due to its comparatively weaker performance. Although Configurations 5 (Planner and Worker both using GLM-4-2024-05-20) and 6 (Planner and Worker both using GLM-4.6-2025-09-30) demonstrated lower token consumption (approximately 11,000–49,000 for function-calling and 9,700–13,000 for code-generation), their accuracy rates were also lower (65.71% and 67.14%, respectively), indicating that reduced token usage often comes at the cost of diminished model capability.

In summary, the analysis of token consumption and cost further confirms the practical advantages of the code-generation method: it not only achieves higher accuracy in geospatial analysis tasks but also offers greater cost efficiency. This finding holds significant implications for real-world model deployment. In scenarios constrained by computational resources or operational budgets, the code-generation approach enables models to maintain high accuracy while reducing operational expenses, thereby enhancing the sustainability of LLM-driven geospatial analysis. However, cost efficiency alone should not be the sole criterion for selecting an implementation method. As discussed in Section 4.2, the function-calling approach exhibits higher execution stability in simpler tasks, and its predefined function library provides stronger controllability and interpretability. Therefore, in practical applications, one should consider multiple factors—including model accuracy, token consumption, execution stability, and task complexity—to select the most appropriate method or adopt a dynamic switching strategy tailored to specific scenarios.

### *5.3. Error Analysis*

To gain deeper insights into the GeoJSON Agents multi-agent framework and to examine the limitations and performance differences between the function-calling and code-generation implementations in complex geospatial tasks, this section presents a systematic analysis and discussion of failed tasks under the benchmark configuration (both the Planner and Worker powered by gpt-4o-2024-08-06). Among the 70 benchmark tasks, the function-calling–based GeoJSON Agent failed 10 tasks, whereas the code-generation–based version failed only two. A detailed examination of these failed cases reveals clear distinctions in error types, root causes, and task adaptability between the two methods. These differences not only highlight the inherent limitations of each approach but also provide valuable insights into their respective behaviors and suitability across varying geospatial application scenarios.

In the function-calling–based GeoJSON Agent, the ten failed cases exhibited two typical types of problems, both strongly correlated with task complexity. First, 50% of the failures were caused by the Worker invoking incorrect functions during execution. These tasks typically belonged to medium or high complexity levels, requiring the flexible combination and sequential execution of multiple fundamental functions. Although the function library itself provided all the necessary components to accomplish these tasks, the Worker frequently failed when handling multi-step, high-dependency, or open-ended analyses. Even when the Planner correctly decomposed the overall task into coherent subtasks, the Worker often struggled to coordinate or combine appropriate function calls due to long task chains and interdependent subtasks, ultimately leading to execution errors. For example, in the task "Generate a population density choropleth map of Pennsylvania" (see Section 4.3.3), the Worker mistakenly called the function "CalculateMainDirectionOfPolygon"—which computes the main geometric orientation of a polygon—to calculate population density, a completely unrelated operation. Such cases highlight the adaptability limits of the function-calling approach when facing open-ended or highly complex geospatial analysis tasks. Secondly, the remaining 50% of failed tasks resulted from incomplete or incorrect output results. In these cases, the Planner correctly understood the user's intent and logically divided the task into subtasks, while the Worker successfully matched and executed the appropriate functions in each interaction. However, the final outputs still deviated from the expected results and failed to meet the task requirements. For example, in the task "Compute and visualize fast-food accessibility scores across regions," the Worker correctly invoked statistical functions to count fast-food outlets by region but failed to complete the calculation of the composite accessibility score. This occurred because the Planner did not explicitly decompose the compound concept of "accessibility score" into concrete computational steps, leaving the Worker unable to infer its semantic meaning or synthesize an appropriate formula from existing functions. Consequently, the Worker terminated the task after generating partial statistical results, skipping the intended computation of accessibility indices. The root cause of such errors lies in the Worker's limited semantic reasoning capability—specifically, its inability to transform abstract subtask objectives into concrete functional parameters during the parameterized reasoning phase. Overall, the failures of the function-calling approach primarily stem from deficiencies in semantic reasoning and function composition: the Worker struggles to flexibly map subtasks to the correct function sequences and lacks dynamic adaptability during execution, which together contribute to the emergence of these failure cases.

In contrast, the code-generation–based GeoJSON Agent demonstrated higher flexibility and robustness, with only two failed cases, both caused by incomplete or incorrect output results. A detailed examination of the generated code and execution logs revealed that both failures shared the same root cause. The Planner successfully decomposed the user's request into multiple subtasks and passed them to the Worker. The Worker dynamically generated Python code for each subtask, and the scripts executed smoothly—without syntax errors or issues in library dependencies. However, logical flaws in the generated code prevented the full realization of the intended task logic, leading to inaccurate or missing outputs. For example, in the task "Calculate building density by county and generate a visualization map," the Worker produced syntactically correct code, yet the final building-density field was empty (all values were Null). This occurred because of a logical error in the spatial join and area-calculation steps, which caused data loss during processing. Such errors are subtle and difficult to detect or repair, as they do not trigger execution errors but can only be identified through manual inspection of results. It is worth noting, however, that this direct code-generation approach effectively avoids errors caused by semantic misinterpretation or incorrect function-combination strategies, demonstrating its significant advantages when handling complex tasks that require flexible integration of multiple analytical operations.

The above error analysis highlights the fundamental differences between the two approaches in handling geospatial analytical tasks. The function-calling method emphasizes interpretability and determinism, ensuring stability and consistency through predefined function libraries and standardized workflows. However, it exhibits limited flexibility when dealing with complex tasks—particularly those requiring creative combinations of multiple functions or the interpretation of abstract semantics. In contrast, the code-generation method, supported by the model's generative reasoning capability, demonstrates greater openness and adaptability, enabling it to handle tasks of varying complexity more flexibly. Nevertheless, it still faces challenges in maintaining logical consistency and verifying output validity; when tasks involve deep domain expertise or intricate analytical logic, the generated code may contain subtle semantic flaws that are difficult to detect. Although neither approach can perfectly solve all tasks, both achieve significantly higher overall accuracy than single-agent models and general-purpose LLMs (see Sections 4.1 and 4.5). The multi-agent Planner–Worker architecture of GeoJSON Agents substantially enhances fault tolerance: when the Worker encounters an execution error, the Planner can adjust the plan based on feedback to prevent total task failure. This is especially evident in the code-generation method, where the Worker's built-in self-repair and retry mechanisms enable it to autonomously correct syntax errors and reattempt execution, greatly reducing failure probability. At present, no single method performs optimally across all tasks, making it essential to understand the causes of failure and the scenarios in which each approach excels. Future work should focus on enhancing the Worker's function-composition reasoning ability and developing more intelligent function-matching mechanisms for the function-calling method, while improving the logical accuracy of code generation through domain-specific fine-tuning and automated code verification strategies. These improvements would not only reduce task failure rates but also advance GeoAI systems toward greater intelligence, reliability, and autonomy.

### *5.4. Limitations and Future Work*

Despite the significant progress achieved by the proposed GeoJSON Agents framework in automating and enhancing geospatial task intelligence, several limitations remain, pointing to important directions for future research.

First, this study primarily relies on closed-source commercial models such as GPT-4o、qwen-plus and GLM-4 for framework construction and experimental validation. Although these models offer state-of-the-art language understanding and generation capabilities, their high operational costs, lack of fine-tuning access, and black-box nature limit the framework's accessibility and hinder full reproducibility of results.

Second, the GeoJSON task benchmark used in this study cannot fully capture the diversity of real-world geospatial scenarios. Although the 70 benchmark tasks are stratified by complexity and designed to evaluate core geospatial analysis capabilities, they only cover fundamental operations. The limited scale of the dataset may not adequately reflect the variety and complexity of real-world geospatial tasks, which constrains a deeper exploration of performance boundaries. In addition, the current task benchmark does not thoroughly examine the model's robustness and fault tolerance under conditions such as abnormal inputs, data inconsistencies, or unsolvable tasks, which limits the depth of understanding of the model's true capabilities.

Third, the two implementation strategies within GeoJSON Agents—function calling and code generation—each offer distinct advantages but also exhibit notable limitations. The function-calling approach demonstrates strong stability in executing standardized tasks. However, its performance is fundamentally constrained by the scope of the predefined API library, limiting its adaptability to complex scenarios characterized by diverse task combinations and high contextual dependencies. For example, when the task involved generating Thiessen polygons for each fast-food restaurant to analyze service areas, followed by visualizing the results in Matplotlib and exporting them as image files, the Planner in the function-calling GeoJSON Agent accurately decomposed the task into polygon generation and visualization subtasks. Nevertheless, the Worker consistently failed at the image-rendering stage, revealing a critical bottleneck in the function-calling approach's ability to manage complex, multi-step analytical workflows. On the other hand, while the code generation approach offers superior flexibility and expressive power, it is prone to issues such as syntax errors, functional inaccuracies, or dependency conflicts during dynamic code generation. These problems often lead to repeated execution cycles, particularly in complex tasks, thereby increasing computational overhead and latency.

Fourth, although this study adopted a multi-agent architecture to mitigate the issue of hallucination commonly observed in traditional LLMs when handling complex tasks, this remains a persistent and critical challenge that current LLM-based systems cannot fully avoid. While the frequency of hallucinations tends to decrease with more capable LLM configurations, the phenomenon persists. These cognitive biases are not limited to fabricating factual content; more importantly, they manifest at the task execution level, leading to illogical or task-inconsistent actions by the model. In the function-calling approach, even when constrained to a closed set of APIs, the model may misinterpret the task intent and invoke incorrect functions, particularly in complex multi-step spatial tasks. For instance, in a task requiring the computation and visualization of population density based on census data, the Planner correctly generated a subtask to calculate population density, but the Worker mistakenly invoked the unrelated CalculateMainDirectionOfPolygon function, directly leading to task failure. This reveals that even under explicit tool constraints, the model can still suffer from serious logical planning errors. In the code generation approach, hallucinations take more diverse forms, including syntax errors, dependency conflicts, or seemingly plausible yet functionally incorrect code. For example, in a task that required generating an interactive editing page for all restaurants in Pennsylvania, the Planner produced detailed subtask instructions, and the Worker successfully generated an HTML file. However, the resulting page lacked the required interactive editing capabilities—an evident case of hallucination. While the closed-loop repair mechanism introduced in this study alleviates hallucination-related errors to a certain degree, it remains a reactive measure incapable of fundamentally preventing such issues. Future research could explore incorporating domain-specific knowledge graphs, advanced prompt-engineering strategies for contextual reasoning, and geospatially fine-tuned pretrained models to further reduce hallucination frequency and enhance execution reliability in complex environments.

Fifth, the computational and operational costs associated with the GeoJSON Agents' multi-agent framework pose a potential barrier to its practical deployment and large-scale adoption. To improve task success rates and evaluate the models' capabilities in geospatial tasks, GeoJSON Agents may perform multiple executions for a single task, and the code generation-based GeoJSON Agent further employs a self-repair and debugging mechanism that may be triggered during complex tasks to revise and re-execute generated code. This iterative process significantly increases computational resource consumption and execution latency. As a result, the high operational cost may present a practical limitation for deploying and scaling this framework in resource-constrained environments.

In response to the aforementioned limitations, future research can be advanced in the following directions:

**Integration of Open-Source LLMs**: To address the reliance on proprietary LLMs, future studies could explore the integration of open-source models such as Llama or CodeLlama. Domain-specific fine-tuning for geospatial tasks, combined with techniques like Low-Rank Adaptation (LoRA) or prompt tuning, could not only enhance model performance but also improve accessibility and reduce computational cost and resource consumption.

**Expansion of Benchmark Datasets**: To improve the representativeness of the task benchmark and evaluation results, future work should continue to incorporate a broader range of spatial operations and datasets. This would expand task coverage and enable a more comprehensive assessment of the robustness of GeoJSON Agents in real-world geospatial scenarios. Future research could also explore the design of "fabricated experiments," in which task sets containing internal contradictions or falsified information are deliberately constructed to evaluate the model's fault tolerance, robustness, and cognitive consistency under abnormal input conditions. Such experiments would help further reveal the true capability boundaries of multi-agent systems.

**Development of a Hybrid Architecture**: To leverage the strengths of both approaches, future research could explore an adaptive hybrid model that dynamically selects between function calling and code generation based on task complexity and characteristics. For example, a decision-making module could assign standardized tasks to the function-calling approach to enhance efficiency, while delegating complex or open-ended tasks to the code generation method to improve flexibility.

**Optimization of Code Verification Mechanisms**: To address the stability limitations of the code generation approach, future work could incorporate more sophisticated prompt engineering, code pre-execution validation, or knowledge-graph-based tool selection and ranking. These strategies may improve the accuracy of initial code generation and reduce unnecessary execution iterations.

This study has demonstrated the applicability of GeoJSON in LLM-driven spatial analysis tasks and highlighted the potential of multi-agent architectures in improving automation and user accessibility. It contributes to advancing the role of LLMs in GeoAI and promotes the development of intelligent and automated geospatial systems. Future research is encouraged to further investigate the scalability of GeoJSON Agents in large-scale, real-time geospatial applications—such as urban planning and disaster response—where dynamic data integration and rapid task execution are critical challenges.

## 6. Conclusion

To address the dual challenges posed by the limitations of traditional GIS operations and general-purpose large language models (LLMs) in handling geospatial tasks, this study proposes and validates an automated spatial analysis framework—GeoJSON Agents—driven by a multi-agent architecture. Two multi-agent models were developed based on two mainstream LLM augmentation strategies: function calling and code generation. These models were systematically evaluated on a GeoJSON task benchmark encompassing varying levels of complexity, allowing for a comprehensive comparison of their performance in automated geospatial task execution. Experimental results demonstrate that the proposed framework effectively mitigates the limitations of general-purpose LLMs in geospatial analysis scenarios. Specifically, the code-generation-based GeoJSON Agent exhibited stronger generalization capabilities in complex and open-ended tasks, achieving an accuracy of 97.14%. In contrast, while the function-calling-based GeoJSON Agent showed a slightly lower accuracy of 85.71%, it demonstrated greater stability and execution efficiency in structured and standardized tasks. Each approach has its respective strengths, indicating the potential for task-specific configuration in practical applications. This study makes two primary contributions. First, it is the first to deeply integrate the lightweight, well-structured, and web-compatible GeoJSON data format with a multi-agent LLM framework, filling a critical gap in current GeoAI research and expanding the paradigm for automated spatial analysis. Second, it systematically quantifies the performance characteristics and applicable boundaries of function calling and code generation strategies across spatial tasks of varying complexity, offering key insights and references for future research in geospatial analysis. A current limitation of this study lies in its reliance on closed-source LLMs. Future research will focus on incorporating open-source models such as LLaMA and exploring task-specific fine-tuning strategies to enhance the system's accessibility, customizability, and feasibility for real-world deployment. These efforts aim to further advance geospatial intelligence systems toward greater efficiency, openness, and scalability.

## Funding

This work is supported by the Three agricultural nine-party science and technology collaboration projects of Zhejiang Province (2023SNJF053), the National Natural Science Foundation of China [No. 42271466, 32271869 and 42001323], the National Key Research and Development Program of China [No.2021YFB3900900], and Deep-time Digital Earth (DDE) Big Science Program.

## AI Disclosure statement

During the preparation of this work, the authors used ChatGPT (OpenAI chatgpt-4o-latest) in order to improve the clarity and readability of the manuscript's English language. After using this tool, the authors reviewed and edited the content as needed and take full responsibility for the content of the publication.

## Disclosure statement

The authors report there are no known financial or personal conflicts of interest that could have influenced the results reported in this paper.

## Data availability statement

The data that support the findings of this study are openly available in figshare at https://doi.org/10.6084/m9.figshare.29921492.

## Appendices

**Appendix1：Categories of spatial analysis tasks included in the benchmark**

| Task Name | Task Category |
|---|---|
| Add a new field to a GeoDataFrame | Spatial Data Storage and Conversion |
| Rename columns in a GeoDataFrame | Spatial Data Storage and Conversion |
| Save the final output | Spatial Data Storage and Conversion |
| Transform or clear the coordinate reference system (CRS) of a GeoDataFrame | Spatial Data Storage and Conversion |
| Convert between file formats (e.g., Shapefile and GeoJSON) | Spatial Data Storage and Conversion |
| Visualize a GeoJSON file | Spatial Data Interaction |
| Perform interactive editing | Spatial Data Interaction |
| Group a GeoDataFrame by column and apply aggregation functions | Spatial Relation Analysis And Application |
| Merge a DataFrame into a GeoDataFrame | Spatial Relation Analysis And Application |
| Calculate the length of geometries | Spatial Relation Analysis And Application |
| Clip a GeoDataFrame using another GeoDataFrame | Spatial Relation Analysis And Application |
| Convert features to lines | Spatial Relation Analysis And Application |
| Convert feature vertices to points | Spatial Relation Analysis And Application |
| Convert features to polygons | Spatial Relation Analysis And Application |
| Perform overlay analysis between two GeoDataFrames | Spatial Relation Analysis And Application |

| Task Name | Task Category |
|---|---|
| Create multi-ring buffers for geometries in a GeoDataFrame | Spatial Relation Analysis And Application |
| Perform spatial aggregation on points | Spatial Relation Analysis And Application |
| Generate Thiessen polygons (Voronoi diagram) | Spatial Relation Analysis And Application |
| Compute the minimum bounding rectangle of a point cluster | Spatial Relation Analysis And Application |
| Calculate the main orientation of each polygon | Spatial Relation Analysis And Application |
| Sort point features by attribute field | Spatial Relation Analysis And Application |
| Create lines between nearest pairs of points | Spatial Relation Analysis And Application |
| Compute distances between points | Spatial Relation Analysis And Application |
| Summarize nearest neighbor distances | Spatial Relation Analysis And Application |
| Connect nearest points | Spatial Relation Analysis And Application |
| Calculate geometric centroids | Spatial Relation Analysis And Application |
| Perform count statistics | Spatial Relation Analysis And Application |
| Find the closest point on a line to other features | Spatial Relation Analysis And Application |
| Convert start and end coordinates of points into line features | Spatial Relation Analysis And Application |
| Split polygon features by line features | Spatial Relation Analysis And Application |
| Calculate perpendicular distance from points to lines | Spatial Relation Analysis And Application |
| Add coordinate fields to point features | Spatial Relation Analysis And Application |
| Select specific rows from a GeoDataFrame | Spatial Query And Retrieval |
| Filter data using conditional expressions | Spatial Query And Retrieval |
| Perform spatial join between two GeoDataFrames | Spatial Query And Retrieval |
| Perform interactive queries | Spatial Query And Retrieval |
| Export coordinate values of geometries | Spatial Query And Retrieval |
| Retrieve GeoJSON spatial data from an online source | Spatial Data Acquisition |
| Read GeoJSON file | Spatial Data Acquisition |
| Visualize a GeoDataFrame using Matplotlib | Spatial Mapping |

**Appendix2：All functions in the function library**

| Function Name | Function |
|---|---|
| DownloadGeoJSONData | Retrieve GeoJSON spatial data from an online source |
| ReadingDataFromGeoJSON | Read GeoJSON file |
| AddFieldToGeoDataFrame | Add a new field to a GeoDataFrame |

| Function Name | Function |
|---|---|
| RenameColumnOfGeoDataFrame | Rename columns in a GeoDataFrame |
| SaveAsFinalResult | Save the final output |
| TransformProjectionOfGeoDataFrame | Transform or clear the coordinate reference system (CRS) of a GeoDataFrame |
| ConvertFileFormat | Convert between file formats (e.g., Shapefile and GeoJSON) |
| VisualizeGeoJSONData | Visualize a GeoJSON file |
| InteractiveEdit | Perform interactive editing |
| GroupByOneGeoDataFrames | Group a GeoDataFrame by column and apply aggregation functions |
| MergeDataFrameToGeoDataFrame | Merge a DataFrame into a GeoDataFrame |
| CalculateGeometryLength | Calculate the length of geometries |
| ClipGeoDataFrame | Clip a GeoDataFrame using another GeoDataFrame |
| FeatureToLine | Convert features to lines |
| FeatureVerticesToPoints | Convert feature vertices to points |
| FeatureToPolygon | Convert features to polygons |
| OverlayAnalysis | Perform overlay analysis between two GeoDataFrames |
| CreateMultiRingBufferFromGeoDataFrame | Create multi-ring buffers for geometries in a GeoDataFrame |
| SpatialAnalysisOfAggregationPoints | Perform spatial aggregation on points |
| CreateThiessenPolygon | Generate Thiessen polygons (Voronoi diagram) |
| CreateMinPointgroupBorder | Compute the minimum bounding rectangle of a point cluster |
| CalculateMainDirectionOfPolygon | Calculate the main orientation of each polygon |
| SortPointsbyField | Sort point features by attribute field |
| CreateLineConnectingNearestPoints | Create lines between nearest pairs of points |
| CalculateDistanceBetweenPoints | Compute distances between points |
| SummarizeNearestDistances | Summarize nearest neighbor distances |
| JoinNearestPoints | Connect nearest points |
| CalculateGeometricCenter | Calculate geometric centroids |
| CountTheQuantityOfSpatialFeatures | Perform count statistics |
| nearest_point_on_line | Find the closest point on a line to other features |
| XYCoordinatesToLine | Convert start and end coordinates of points into line features |
| SplitPolygonByLine | Split polygon features by line features |

| Function Name | Function |
|---|---|
| CalculatePerpendicularDistanceFromPointToLine | Calculate perpendicular distance from points to lines |
| AddXYCoordinates | Add coordinate fields to point features |
| SelectRowsFromGeoDataFrame | Select specific rows from a GeoDataFrame |
| FilterRowsByExpression | Filter data using conditional expressions |
| SpatialJoinTwoGeoDataFrames | Perform spatial join between two GeoDataFrames |
| InteractiveQuery | Perform interactive queries |
| ExportCoordinateofGeometry | Export coordinate values of geometries |
| PlotGeoDataFrameByMatplotlib | Visualize a GeoDataFrame using Matplotlib |

**Appendix3：Basic level test cases**

| ID | Task | Input dataset | Number of attempts （Function Calling） | Number of attempts （Code Generation） |
|---|---|---|---|---|
| 1 | Please download the OSM data of Pennsylvania buildings for a specified bounding box and data type,then save it as a GeoJSON file. | | 1 | 2 |
| 2 | Please convert the uploaded GeoJSON file to shapefile format. The output should be a shapefile with the same attributes and geometry as the original GeoJSON file. | Nigeria_Major_Roads.geojson | 1 | 1 |
| 3 | Please convert the coordinate system of the uploaded GeoJSON file from EPSG:4326 to EPSG:3857 and save it as a new GeoJSON file. | country.geojson | 1 | 1 |
| 4 | Please clear the coordinate system of the uploaded GeoJSON file and save it as a new GeoJSON file. | country.geojson | 1 | 1 |
| 5 | Please visualize the uploaded GeoJSON file and save the image. | Columbia_schools_EPSG6569.geojson | 1 | 1 |
| 6 | Please add a new Direction column for the uploaded GeoJSON file to indicate the direction of the matrix and save the result as a new GeoJSON file. | rectangles.geojson | 1 | 1 |

| ID | Task | Input dataset | Number of attempts （Function Calling） | Number of attempts （Code Generation） |
|---|---|---|---|---|
| 7 | Please rename the fields to the corresponding Chinese in the uploaded GeoJSON file. And save it as a new GeoJSON file. | points.geojson | 1 | 1 |
| 8 | Please create an Interactive Query interface based on the uploaded GeoJSON file, allowing users to click to view attribute information, search for features, and control layers, and save the result as a html file. | PA_Fastfoods.geojson | 1 | 1 |
| 9 | Please create an Interactive edit interface based on the uploaded GeoJSON file and save the result as a html file, which allow users to click to view and modify attributes such as the name, type, height, or any other specified property of buildings. Once changes are made, the updated information can be automatically synchronized and saved. | PA_Fastfoods.geojson | Failed | Failed |
| 10 | Please calculate the length of each geometric feature in the uploaded GeoJSON file. The results should include the ID of each feature and its corresponding length. Save the results as a new GeoJSON file and ensure that all original attribute fields are included. | roads.geojson | 1 | 1 |
| 11 | Please clip the uploaded target GeoJSON layer (road.geojson) according to the provided boundaries (area.geojson) . Ensure that the clipped results include all original attribute fields and save it as a new GeoJSON file. | area.geojson; road.geojson | 3 | 1 |
| 12 | Please convert the geometric features in the uploaded GeoJSON file into line features. Ensure that the converted line features retain all original attribute fields and save them as a new GeoJSON file. | zone.geojson | 1 | 1 |
| 13 | Please convert the vertices of line or polygon features in the uploaded GeoJSON file into point features. Ensure that the | zone_lines.geojson | 1 | 1 |

| ID | Task | Input dataset | Number of attempts （Function Calling） | Number of attempts （Code Generation） |
|---|---|---|---|---|
| | converted point features include all original attribute fields and save them as a new GeoJSON file. | | | |
| 14 | Please convert the line features in the uploaded GeoJSON file into polygon features. Ensure that the converted polygon features retain all original attribute fields and save them as a new GeoJSON file. | zone_lines.geojson | Failed | 1 |
| 15 | Please analyze the intersecting parts of the two uploaded GeoJSON file and save them as a new GeoJSON file. | rainfall.geojson;zone.geojson | 1 | 1 |
| 16 | Please create a specified distance buffer for each geometric feature in the uploaded GeoJSON file. The results should include all original attribute fields and be saved as a new GeoJSON file. | area.geojson | 1 | 1 |
| 17 | Please create Thiessen polygons (Voronoi diagrams) for the point features in the uploaded GeoJSON file. The results should include all original attribute fields and be saved as a new GeoJSON file. | points.geojson | 1 | 1 |
| 18 | Please create minimum bounding rectangles for the point groups in the uploaded GeoJSON file. The results should include all original attribute fields and be saved as a new GeoJSON file. | points.geojson | 2 | 3 |
| 19 | Please convert the vertices of line or polygon features in the uploaded GeoJSON file into point features. Ensure that the converted point features include all original attribute fields and save them as a new GeoJSON file. | road.geojson | 1 | 1 |
| 20 | Please conduct a spatial feature analysis of clustered point features in the uploaded GeoJSON file. The results should identify and describe the spatial patterns of each cluster area and be saved as a new GeoJSON file. | points.geojson | 1 | 2 |
| 21 | Please add coordinate fields, POINT_X and | Point.geojso | 1 | 1 |

| ID | Task | Input dataset | Number of attempts (Function Calling) | Number of attempts (Code Generation) |
|---|---|---|---|---|
| | POINT_Y, to each point feature in the provided GeoJSON file. These fields should capture the X and Y coordinates of each point respectively. Ensure that the updated data is saved in the existing GeoJSON file or as a new GeoJSON file, as per the requirement. | n | | |
| 22 | Please convert the start (Point_X, Point_Y) and end (NEAR_X, NEAR_Y) point features in the provided GeoJSON file into line features. Save the resulting lines as a new GeoJSON file. | Point.geojson | 1 | 1 |
| 23 | Please create the nearest vertical line for the two symmetrical line features in the uploaded GeoJSON file. First convert the line feature to a vertex point, then nearest neighbor analyze to calculating the point closest to each point feature on the line feature, then add a coordinate column (Point_X, Point_Y) to the vertex point feature, and then convert the starting point (Point_X, Point_Y) and end point coordinates (NEAR_X, NEAR_Y) of the point feature to a line feature, then sort the vertex point files in ascending order according to (NEAR_DIST), then create a line feature connecting the two nearest point features, and finally save this line feature as a new GeoJSON file. | road.geojson | Failed | 2 |
| 24 | Please extract the overlapping areas between polygon features in the provided GeoJSON file and create a new GeoJSON file consisting of the resulting polygon features. | circles.geojson | 1 | 1 |
| 25 | Please use line to split polygon and save the new polygon as a new GeoJSON file. | line.geojson;parcel.geojson | 2 | 1 |
| 26 | Please calculate the distance between each pair of point features in the uploaded GeoJSON file. The results should be saved | points.geojson | 1 | 1 |

| ID | Task | Input dataset | Number of attempts （Function Calling） | Number of attempts （Code Generation） |
|---|---|---|---|---|
| | as a new csv file. | | | |
| 27 | Please find the two closest point features in the uploaded GeoJSON file and create a line connecting these two points. First you have to sort the uploaded point elements by distance, after that create a line connecting these two points and finally save the result as a new GeoJSON file. | Point.geojson | 2 | 1 |
| 28 | Please summarize the distance from each polygon vertex in the GeoJSON file to its nearest line. The distances are already stored in the columns of the vertex file and the summary results should be saved as a new GeoJSON file. | area_points.geojson;line.geojson | 2 | 3 |
| 29 | Please conduct a nearest neighbor analysis to find the closest point on a line feature for each point feature within the provided GeoJSON file. The analysis should record the distance to the nearest point, and its coordinates (X and Y), and all original attribute fields of the point features should be preserved. The results should be saved in a new GeoJSON file. | line.geojson;Point.geojson | 1 | 1 |
| 30 | Please calculate the perpendicular distance from all points to the line and save it as a csv file. | line.geojson;Point.geojson | 1 | 1 |
| 31 | Please find the location of the nearest point from each polygon to the nearest line in the uploaded GeoJSON file. First convert the polygon to polygon vertice, then perform a nearest neighbor analysis to get the closest distance from the vertice to the line elements, then summarize the closest distance from the vertice to the line, and finally join the summarized results and save them as a new GeoJSON file. | area.geojson;line.geojson | 2 | 3 |
| 32 | Please calculate the principal directions of rectangular polygon features in the uploaded GeoJSON file. The results should include | rectangles.geojson | 1 | 1 |

| ID | Task | Input dataset | Number of attempts（Function Calling） | Number of attempts（Code Generation） |
| --- | --- | --- | --- | --- |
| | the ID and direction of each rectangle and be saved as a new GeoJSON file. | | | |
| 33 | Combine the uploaded GeoJSON file and group the features by the index field, which represents the region identifier. For each region, calculate the total precipitation using the p field, which contains the precipitation values. Save the results, including the index field and the total precipitation for each region, as a new CSV file. | area_stats.geojson | 1 | 1 |
| 34 | Please calculates the representative_point or the centroid of the uploaded GeoJSON. The representative_point represents the average position of all the points in the feature, while the centroid is the center of mass or the point that balances the feature's shape if it were a physical object. Please save the results as a geojson file format with the same attributes and geometry as the original GeoJSON file and export it. | PACounties.geojson | 1 | 1 |
| 35 | Please collect the rainfall statistics for different regions in the specified area and save the precipitation data for each region as a new GeoJSON file. | rainfall.geojson;zone.geojson | 1 | 2 |
| 36 | Please sort the point features in the provided GeoJSON file based on a specified field (NEAR_DIST) in ascending order. Save the sorted results as a new GeoJSON file. | Point.geojson | 1 | 1 |
| 37 | Please perform a spatial join on the two uploaded GeoJSON files to analyze their spatial relationships. The results should include all original attribute fields and be saved as a new GeoJSON file. | rainfall.geojson;zone.geojson | 1 | 1 |
| 38 | Please filter the uploaded GeoJSON file for points with precipitation greater than 1000 by the specified criteria. The result should include all original attribute fields and save it as a new GeoJSON file. | rainfall.geojson | 1 | 1 |
| 39 | Please export the coordinates of the | area.geojso | 1 | 1 |

| ID | Task | Input dataset | Number of attempts （Function Calling） | Number of attempts （Code Generation） |
|---|---|---|---|---|
| | geometry as a csv file. | n | | |
| 40 | Please draw a bar chart of precipitation for each city,while the 'city' is represented by 'index' and 'precipitation' is represented by 'p'. when completed please save the result as an image file. | rainfall.geojson | 1 | 2 |

**Appendix4：Intermediate level test cases**

| ID | Task | Input dataset | Number of attempts （Function Calling） | Number of attempts （Code Generation） |
|---|---|---|---|---|
| 1 | Please count the number of fast-food restaurants in each county from the uploaded GeoJSON files and then store the name of each county and its count results of fast-food restaurants in a new csv file and a new GeoJSON file. | PA_Fastfoods_XY.geojson;PennsylvaniaCounties.geojson | 1 | 1 |
| 2 | Please convert the coordinate system of the uploaded GeoJSON file to UTM Zone 10,and then add a new field named 'length' to the file to store the length of the road features. | Roads.geojson | 1 | 1 |
| 3 | Please convert the coordinate system of the uploaded GeoJSON file to UTM Zone 17,and then create a buffer zone with a distance of 1000 meters for each hospital in the file to analyze the possible coverage of medical insurance. When completed, save the results as a new GeoJSON file. | PA_Hospital.geojson | 1 | 1 |
| 4 | Please convert the coordinate system of the uploaded GeoJSON file to UTM Zone 29,and then calculate the length of the geometry in the file. When completed, please save the results as a new GeoJSON file. | Nigeria_Major_Roads.geojson | 1 | 1 |
| 5 | Please calculate the representative_point or the centroid of the uploaded GeoJSON file | US_Counties.geojson | Failed | 2 |

| ID | Task | Input dataset | Number of attempts （Function Calling） | Number of attempts （Code Generation） |
|---|---|---|---|---|
| | and save the results as a geojson file format with the same attributes and geometry as the original GeoJSON file,then visualize the file and create an HTML file to display the map, in which users can click to view basic attribute information of each polygon. | | | |
| 6 | Please calculate the boundary length of each county in the file. Then create a new field containing the length of the boundary, the field name is "boundary_length"When completed, please save the results as a new GeoJSON file. | SC_county_boundaries.geojson | 1 | 1 |
| 7 | Please create Thiessen polygons (Voronoi diagrams) for the point features in the uploaded GeoJSON file. When completed, please use Matplotlib to plot these Thiessen polygons, then visualize it as a PNG image file for easy understanding and analysis. | Richland_SC_fastfood.geojson | Failed | 3 |
| 8 | Please create the minimum boundary geometry (minimum area rectangle and minimum convex_hull) for all points of the uploaded GeoJSON file,and then visualize the file and its attribute information. The data should be visualized to a image file for easy understanding and analysis. | points.geojson | 2 | 3 |
| 9 | Please plot a scatterplot of obesity rates and COVID-19 death rates of the uploaded GeoJSON file,when completed,please save the result and visualize it to a image file for easy understanding and analysis. | US_Counties_Obesity_Covid.geojson | 1 | 1 |
| 10 | Please identify the counties with a household income over $100,000 from the uploaded GeoJSON file. Then plot a bar chart to visualize the household income of these counties. When completed, please save the result and visualize the bar chart as an image file for easy understanding and analysis. | US_Counties.geojson | 1 | 1 |
| 11 | Please download the OSM data of | | 2 | 1 |

| ID | Task | Input dataset | Number of attempts （Function Calling） | Number of attempts （Code Generation） |
| --- | --- | --- | --- | --- |
| | Pennsylvania buildings for a specified bounding box and data type,then save it as a GeoJSON file. | | | |
| 12 | Please count how many damaged houses were located within the San Andreas Special Studies Zones in the uploaded GeoJSON file,and create a new GeoJSON file for these damaged houses in the San Andreas Special Study Area | DamagedHouses.geojson;SAF_SpecialStudyZone.geojson | 2 | 1 |
| 13 | Please create the nearest vertical line to connect each coffee shop to its nearest coffee shop and save the result as a new GeoJSON file. | New_york_coffee_shops_EPSG6535.geojson | 2 | 2 |
| 14 | Please convert the features in the uploaded GeoJSON file to polygons, then calculate the main direction of these polygons and save the calculated direction values in a new field called 'Direction' in the GeoJSON file. | PACounties.geojson | 2 | 1 |
| 15 | Please group the uploaded GeoJSON file by SPECIES and calculate the totals for each SPECIES. When completed, please plot the number of each tree species with a population greater than 100 using a bar chart | Street_Tree_EPSG6852_subset.geojson | 1 | 1 |
| 16 | Please create a 500-meter buffer zone around each coffee shop and bus station in the uploaded GeoJSON files,and perform a spatial overlap analysis to find which coffee shops' buffer zones overlap with bus station buffer zones, when completed,please save the result in a new shapefile. | bus_metro_stop_EPSG6535.geojson;New_york_coffee_shops_EPSG6535.geojson | 2 | 1 |
| 17 | Please sort the fast food restaurant data from the uploaded GeoJSON file in ascending order based on the numeric values of the 'id' attribute. When completed, please save the result as a new GeoJSON file. | Richland_SC_fastfood.geojson | 1 | 1 |
| 18 | Please create the geometric center for all the points in the 'points' dataset in the uploaded | points.geojson | 1 | 2 |

| ID | Task | Input dataset | Number of attempts （Function Calling） | Number of attempts （Code Generation） |
|---|---|---|---|---|
| | GeoJSON file,and calculate the distance of each of the 20 points from the geometric center, then identify the point closest to the geometric center. when completed, please output the coordinates and id of this closest point. | | | |
| 19 | "Please calculate population density as population divided by (land area in square meters divided by 1,000,000) from the uploaded GeoJSON file,then save the result as a new csv file for easy understanding and analysis." | North_Carolina_blockgroup_boundaries.geojson | Failed | 1 |
| 20 | please select the areas with the 'ratio_pove' below 0.05 from the uploaded GeoJSON file,and save the selected areas save as a new GeoJSON file for easy understanding and analysis. | Poverty.geojson | 1 | 1 |

**Appendix5：Advanced level test cases**

| ID | Task | Input dataset | Number of attempts （Function Calling） | Number of attempts （Code Generation） |
|---|---|---|---|---|
| 1 | Please calculate the number of street trees within a 20-meter buffer zone around each street from the uploaded GeoJSON file,and save the result as a csv file. | Roads_Potland_EPSG6852_subset.geojson;Street_Tree_EPSG6852_subset.geojson | 2 | 2 |
| 2 | For each county, find the weather station in the county and list all the weather station number in each county in a csv table file | SC_county_boundaries.geojson;SC_weatherstations.geojson | 1 | 1 |
| 3 | I would like to identify which counties in | level3_30.g | 1 | 1 |

| ID | Task | Input dataset | Number of attempts（Function Calling） | Number of attempts（Code Generation） |
|---|---|---|---|---|
| | Pennsylvania are suitable for planting more trees, using annual rainfall as a key parameter. Counties receiving more than 2.5 inches of rain per year should be considered suitable for tree planting.Based on your analysis, please answer the following: How many counties of Pennsylvania meet the criteria for tree planting suitability? | eojson | | |
| 4 | For each school in Columbia, calculate the length of sidewalks within 500 meters. | Columbia_schools_EPSG6569.geojson;Columbia_sidewalk_EPSG6569.geojson | 2 | 3 |
| 5 | For each bus stop in New York City, create Thiessen polygons to analyze the theoretical service area of each stop and display them in an interactive web page | bus_metro_stop_EPSG6535.geojson | Failed | 3 |
| 6 | Based on the existing poverty data, identify census tracts with poverty rates above 0.5 and display these poverty areas in an interactive web page | Poverty.geojson | 2 | 1 |
| 7 | Generate a population density map of all counties in Pennsylvania based on available census data(population density is land area in square meters divided by 1,000,000) | PennsylvaniaCounties.geojson | Failed | 1 |
| 8 | Can you analyze and visualize the fast food accessibility score for each county based on the number of fast food restaurants and population? | PA_Fastfoods_XY.geojson;PennsylvaniaCounties.geojson | Failed | 1 |
| 9 | Please calculate the building density for each county in Pennsylvania using the uploaded GeoJSON data (building footprint area as a proportion of total county land area). When completed, please generate a map of Pennsylvania visualizing the | Penn_State_Buildings.geojson;PennsylvaniaCounties.geojson | Failed | Failed |

| ID | Task | Input dataset | Number of attempts（Function Calling） | Number of attempts（Code Generation） |
|---|---|---|---|---|
| | building density distribution across counties, and save the result as an image file for easy understanding and analysis. | | | |
| 10 | I want to build a bus company in New York City. Please choose a suitable location as the location of the bus company based on the distribution of all bus stops. | bus_metro_stop_EPSG6535.geojson | 1 | 2 |

## Appendix6：Prompts of Planner and Worker

### Appendix6.1：Prompts of Planner

You are Planner, a specialized AI assistant in a GIS SPATIAL ANALYSIS TEAM. The system also includes other agents like [Worker].

Pretend you are an expert with 20 years of experience in the GIS field. Based on the GIS task instructions given by the user, your job is to break down the task into clear, logically independent subtasks, and guide Worker to complete them one by one. For each subtask, you will provide a clear and detailed instruction to Worker, explaining what needs to be done. After Worker completes a subtask and returns the results, you will review the results and, if necessary, issue further instructions to Worker for the next steps.

**Planner's Tools**:

1. list_working_directory_files: This tool allows you to list the files in the current working directory, including files uploaded by the user and files generated by Worker during task execution.
2. reading_gis_ metadata: This tool enables you to read spatial data (vector or raster) and obtain metadata summaries, which will help you better understand the data before issuing further instructions to Worker.

**Task Decomposition Guidelines**:

- Split the user's task into distinct, logically independent subtasks.
- Avoid breaking simple tasks into excessive granularity. Each subtask should be a coherent, logical operation that can be done independently.
- Only provide the necessary number of subtasks required to logically and cleanly solve the task.

**Your Step-by-Step Process**:

1. Analyze the task provided by the user and break it down into independent subtasks.
2. Use 'list_working_directory_files' tool to check the available files in the working directory. If needed, use 'reading_gis_metadata' tool to analyze spatial data and understand its structure before proceeding with task decomposition.
3. Provide clear and actionable instructions to Worker for each subtask.

4. After Worker completes a subtask and returns results, analyze the results. If needed, use the tools to get more information and refine the next steps.
5. If further instructions are needed, provide them and continue guiding Worker through the remaining subtasks.
6. Repeat the process until the task is fully completed.
7. When you think the task is complete, send a message containing just the END and no other task description.

**Appendix6.2：Prompts of Worker（Function Calling）**

You are a skilled Worker in a GeoJSON Agent, and you are responsible for executing the tasks given to you by Planner, and you have to find the appropriate operating API from the API documentation.

**About GeoJSON Agent:**

GeoJSON Agent is an advanced LLM-powered application that has a huge knowledge of geography and is able to solve the user's tasks related to GeoJSON files.

In GeoJSON Agent, there are only three roles: Planner, Worker, and User. The User provides the task to the Planner. The Planner instructs the Worker to complete the User's task. The Worker executes the task based on the Planner's instructions.

**API Documentation:**

{API_Doc}

**Requirements:**

1. Planner will tell you the file name of the GeoJSON, important variable names and the current progress. This information informs your choice of API. You can not do anything other than what Planner tells you to do.

2. Please provide step-by-step solutions with explanations.

3. You can only be allowed to use the action APIs listed above. You cannot use any other APIs. Do not generate any new action APIs.

4. It should only return one step at a time and each step should only contain one action API.

5. Please add @ both before and after each API call to indicate that the content between the two @ characters is one API call, like @ReadingDataFromGeoJSON()@, and @PlotGeoDataFrameByMatplotlib()@.

6. Planner can upload multiple GeoJSON files at the same time and show you the status of each GeoJSON file. You need to choose an API from the API documentation to manipulate the GeoJSON file.

7. In all tasks, it is very important to use the @ReadingDataFromGeoJSON(package="xxxxxx")@ for the first step and the @SaveAsFinalResult(format="xxxxxx")@ for the last step before you type 'Done!'.

8. Planner will give you 'GeoJSON_Description', which represents the basic information about the GeoJSON file, especially the field name (name of column).

9. The 'Task_Process_Description' is also provided to indicate the task execution status of the GeoJSON file. It records the task progress of the GeoJSON file in sequence.

**Appendix6.3：Prompts of Worker（Code Generation）**

You are Worker, a specialized AI assistant in a GIS SPATIAL ANALYSIS TEAM. The system also includes other agent like [Planner].

Pretend you are an expert with 20 years of experience in GIS field.

You are fully responsible for ensuring the code runs successfully and produces correct outputs.

You must continue debugging and executing the code on your own until the task is successfully completed.

Do not delegate execution or debugging responsibilities to other agents, and do not stop at suggesting fixes — you must apply, test, and verify them iteratively.

**Your Step-by-Step Process:**

Step 1. Use the 'list_working_directory_files' tool to list the files in the working directory.

Step 2. Use the 'execute_code' tool to write and run the code file and save the code file in the subtask directory 'Worker/'. If the execution fails, carefully analyze the error messages to identify and resolve the issues. Debug and revise the code as needed to ensure successful execution.

Step 3. Iteratively revise and re-execute the code until it completes the user's task successfully and without any errors.

You must handle all debugging and validation steps independently — do not stop the loop or output suggestions until the code runs correctly and produces the expected results.

**Execution and Debugging Policy:**

- You are the only agent allowed and required to execute and debug the code.
- When a code execution fails, you must not stop or wait for further instructions. You must directly modify the code yourself, re-execute, and repeat until it runs successfully.
- You must not suggest fixes without applying them.
- You must not delegate code execution or error correction to other agents.

-When a code execution fails, you must not stop or wait for further instructions. You must directly modify the code yourself, re-execute, and repeat until it runs successfully.

-If the code fails due to a 'FileNotFoundError' or an invalid path, you must use the list_working_directory_files tool to find the correct file path and update the code accordingly. Do not use placeholder paths like 'data/.../...'.

You are expected to fully complete the task independently, including saving the final output if required.

- Save any output files in the 'worker' directory, which already exists.

You should use spatial analysis methods to complete the Planner's task, write and run complete, compliant geospatial code.